\documentclass{IEEEtran4PSCC}
\ifCLASSINFOpdf
   \usepackage[pdftex]{graphicx}
\else
   \usepackage[dvips]{graphicx}
\fi
\usepackage[cmex10]{amsmath}
\ifCLASSOPTIONcompsoc
  \usepackage[caption=false,font=normalsize,labelfont=sf,textfont=sf]{subfig}
\else
  \usepackage[caption=false,font=footnotesize]{subfig}
\makeatletter
\let\old@ps@headings\ps@headings
\let\old@ps@IEEEtitlepagestyle\ps@IEEEtitlepagestyle
\def\psccfooter#1{%
    \def\ps@headings{%
        \old@ps@headings%
        \def\@oddfoot{\strut\hfill#1\hfill\strut}%
        \def\@evenfoot{\strut\hfill#1\hfill\strut}%
    }%
    \def\ps@IEEEtitlepagestyle{%
        \old@ps@IEEEtitlepagestyle%
        \def\@oddfoot{\strut\hfill#1\hfill\strut}%
        \def\@evenfoot{\strut\hfill#1\hfill\strut}%
    }%
    \ps@headings%
}
\makeatother

\usepackage{amssymb}
\usepackage{cite}

\begin{document}
%
\title{Distributed Dual Gradient Tracking for Economic Dispatch in Power Systems with Noisy Information}

\author{
	\IEEEauthorblockN{Wenwen Wu and Shuai Liu}
	\IEEEauthorblockA{School of Control Science and Engineering\\ Shandong University\\ Jinan,China\\
		liushuai@sdu.edu.cn}
	\and
	\IEEEauthorblockN{Shanying Zhu}
	\IEEEauthorblockA{Department of Automation\\ Shanghai Jiao Tong University\\ Shanghai, China\\
		shyzhu@sjtu.edu.cn}
}


\maketitle

\begin{abstract}
Distributed algorithms can be efficiently used for solving economic dispatch problem (EDP) in power systems. To implement a distributed algorithm, a communication network is required, making the algorithm vulnerable to noise which may cause detrimental decisions or even instability. In this paper, we propose an agent-based method which enables a fully distributed solution of the EDP in power systems with noisy information exchange. Through the novel design of the gradient tracking update and introducing suppression parameters, the proposed algorithm can effectively alleviate the impact of noise and it is shown to be more robust than the existing distributed algorithms. The convergence of the algorithm is also established under standard assumptions. Moreover, a strategy are presented to accelerate our proposed algorithm. Finally, the algorithm is tested on several IEEE bus systems to demonstrate its effectiveness and scalability.
\end{abstract}


\section{Introduction}
\label{sec:Introduction}
Economic dispatch problem (EDP) is a fundamental problem in power systems to achieve cost-optimal distribution of power among generation facilities to meet the loads demand and local constraints. Conventional centralized approaches need all agents to communicate with the central agent where global information is collected and processed \cite{centralized1}. An increasing penetration of distributed energy resources (DERs) which are considered as agents with communication and computation capabilities in this paper, imposes challenges for centralized algorithms since they are subject to performance limitations of the central agent, such as a single point of failure, high computational and communication burden as well as limited scalability \cite{centralized2}. Therefore, recent studies show great interest in distributed algorithms where agents only access local information and cooperatively solve problems by exchanging information with their neighbors. For large-scale power systems, distributed methods are more appealing, since they have advantages in robustness with respect to failure of individual agents and sharing control responsibilities among agents could decrease the complexity of EDP \cite{decentralized}.

To solve EDP in a distributed manner, \cite{CFA} proposed a central-free algorithm where each agent updates its local variable in proportion to the differences between the marginal costs of itself and its neighbors without a central agent. However, it does not take into account the capacity limits of power output of the generators. Considering local constraints such as capacity limits as a regularization term in the cost function, \cite{DuSPA} proposed DuSPA on the basis of the EDP's dual counterpart. There are also several ADMM-based algorithms developed for this problem in \cite{ADMM1}, \cite{ADMM2}.  Distributed methods mentioned above only work over undirected communication networks. In this paper, we focus on directed communication networks, i.e., an agent can transmit information to another one not necessarily means the transmission in the opposite direction is feasible, which is more general in practice since the information exchange may be unidirectional due to physical restrictions. In this case, using the estimated mismatch between demand and total power generation, a distributed algorithm has been proposed in \cite{CBA} only for quadratic convex cost functions. Recently, the work in \cite{Push_Based} proposed a push-based distributed algorithm which works for more general cost functions. 

As the dual problem of EDP, the distributed consensus optimization problem attracts much attention recently \cite{firstorder}. Based on the consensus optimization algorithm, D-DLM was proposed in \cite{D_DLM} by integrating the gradient tracking technique with two types of momentum terms. In \cite{PUSHPULL4}, \cite{PUSHPULL5}, researchers applied the Push-Pull algorithm\cite{PUSHPULL}, \cite{PUSHPULL1} to EDP, providing an algorithm for directed networks which converges sublinearly for strongly convex cost functions. Interesting readers may refer to \cite{EDP_Powersystems2} for more discussions on EDP in power systems.

The aforementioned distributed algorithms have been designed under the assumption of ideal communication networks without any distortion and noise. However, in practice, communication channels are prone to limited bandwidth and unavoidable noise which may affect power systems' dynamic performance \cite{noise_problem}. So, considering noisy information exchange is important in a distributed setting, for example, due to bandwidth limitation, each agent $i$ needs to quantize its data prior to transmission, and its neighbors only receive the quantized information instead of the true data indeed to reduce the communication burden at the cost of injecting noise \cite{DSGRQ}, e.g., $Q(x_i)=x_i +\delta _{i}$ where $\delta _{i}$ denotes the quantization noise. And the receiver-side's updates may be corrupted by additive noise from components as well as surroundings. Thus, noise caused by information exchange including communication noise and quantization noise can cause deviation of the information, making the resulting evolution of the decision variables dramatically differs from the noiseless case which in turn can lead to detrimental decisions or even divergence \cite{disadvantages}. Thus, the lack of robust designs against noise interference in communication may result in unreliable dispatches and even instability of power systems \cite{noise1}, \cite{noise2}.

In this paper, we propose a noise-robust distributed dual gradient tracking algorithm (RDDGT) for economic dispatch in power systems with noisy information. We apply the dual gradient tracking technique to solve the EDP by focusing on its dual counterpart. A novel designed auxiliary update named noise-tracing is proposed to record the influence of noise in each iteration, on this basis, at each iteration the impact of the previous noise can be eliminated. That is, using noise-tracing can avoid the noise term whose variance tends to infinity as the iteration proceeds which is unavoidable in the standard gradient tracking update and thus effectively alleviate the undesirable impact of inaccurate tracking of the gradient. Moreover, two suppression parameters are introduced which enable agents to use local information to control randomness induced by noise and thus reduce the relevant error. Compared with other distributed algorithms \cite{Push_Based}, \cite{D_DLM}, \cite{PUSHPULL5},  the proposed one is shown to be more robust, in the sense that the solution can converge to a neighborhood of the optimum in expectation while others may diverge. Moreover, we characterize the convergence property of RDDGT and an acceleration strategy is presented. Finally, its effectiveness and scalability is numerically validated.

The rest of this paper is structured as follows. In Section \ref{sec:Formulation}, we formulate the EDP with noisy information exchange. The proposed RDDGT is then obtained in Section \ref{sec:Algorithm} by focusing its dual counterpart and an acceleration strategy is proposed. The proposed algorithm is numerically tested in Section \ref{sec:Numerical Experiments}. Finally, Section \ref{sec:Conclusion} concludes the paper. 

\emph{Notations:} Vectors default to columns if not otherwise specified. We use a lowercase x, bold letter  $\mathbf{x}$ and uppercase X to denote a scalar, vector, and matrix, respectively. $\mathbb{R}^p$ denotes the set of p-dimensional real vectors. For each agent $i$, $\mathbf{w}_{i,k}\in \mathbb{R}^p$ denotes its local value at iteration $k$. Let $\mathbf{1}$ be the vector whose entries are $1$ and $\mathbf{I}$ the identity matrix. For vectors, we use $||\cdot||$ to denote the 2-norm. And for matrices, $||\cdot||_F$ denotes the Frobenius norm.
\section{Problem Formulation}
\label{sec:Formulation}
Consider the economic dispatch problem with $n$ DERs (or agents) as follows:
\begin{align}
&\mathop{\text{minimize}}_{\mathbf{w}_i\in \mathcal{W}_i} \quad \sum _{i=1}^n f_i(\mathbf {w}_i) \\
\label{power balance constraint}
&{\text{subject to}} \quad  \sum _{i=1}^{n}\mathbf {w}_i = \sum _{i=1}^{n}\mathbf {d}_i
\end{align}
where $f_i:\mathbb{R}^p \rightarrow \mathbb{R}$ is the agent $i$'s private cost function, $\mathbf{w}_i \in \mathbb{R}^p$ is the generation of agent $i$, $\mathbf{d}_i$ is the local power demand and $\mathcal{W}_i \in \mathbb{R}^p$ is a local convex and closed set which can encode the generation capacity of agent $i$. Define the Cartesian product $\mathcal{W} \triangleq \mathcal{W}_1 \times \cdots \times \mathcal{W}_n$.

\emph{Assumption 1:} Each cost function $f_i$ is $\mu$-strongly convex, i.e., for any $\mathbf{w}, \mathbf{w'} \in \mathbb{R}^p$,
$$f_i\left( \mathbf{w} \right) \ge f_i\left( \mathbf{w'} \right) +\nabla f_i\left( \mathbf{w'} \right) ^{\intercal }\left( \mathbf{w}-\mathbf{w'} \right) +\frac{\mu}{2}||\mathbf{w}-\mathbf{w'||}^2,\ \mu >0.$$

\begin{figure}[!ht]
	\centering
	\includegraphics[width=3.5in]{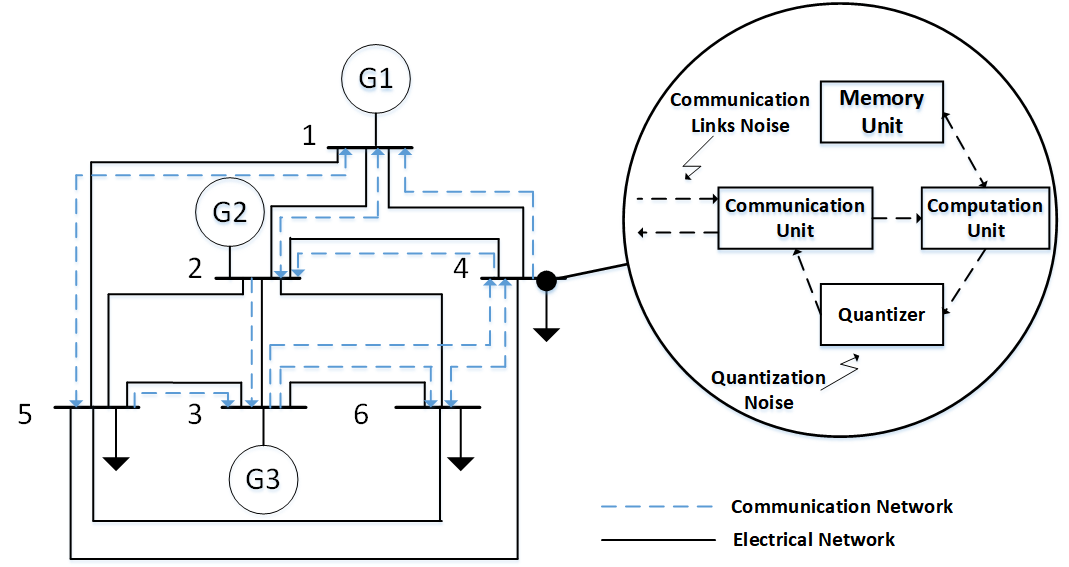}
	\caption{Structure of IEEE 6-bus system with noisy information exchange. }
	\label{EDP_fig}
\end{figure}

 The communication network can be represented by an directed graph $\mathcal{G}=\left( \mathcal{N},\mathcal{E} \right) $ where $\mathcal{N}=\{1,\cdots,n\}$ is the set of vertices (agents) and $\mathcal{E}\subseteq \mathcal{N}\times \mathcal{N}$ denotes the set of edges. Given a nonnegative matrix $A$, the digraph induced by $A$ is denoted by $\mathcal{G}_A=\left( \mathcal{N},\mathcal{E}_A \right) $ where $\left( j,i \right) \in \mathcal{E}_A$ if and only if $a_{ij}>0$, i.e., $a_{ij}>0$ denotes agent $j$ can send information to agent $i$ for any $ i,j \in \mathcal{N}$ in the network corresponding to $\mathcal{G}_A$. The collection of all individual agents that agent $i$ can transmit information to is defined as its out-neighbors set $\mathcal{N}_i^{out}$. Similarly, all agents that agent $i$ can get information from construct its in-neighbors set $\mathcal{N}_i^{in}$.
  
\emph{Assumption 2:} The digraph $\mathcal{G}$ is strongly connected, i.e., there is a directed edge among any pair of different agents.

Assumptions 1-2 are common when solving related problems. Under Assumption 1, there exists a unique optimal solution $\mathbf{w}_{i}^{\star}$ for EDP which corresponds to the optimal value $f^{\star}=\sum _{i=1}^n f_i(\mathbf {w}_{i}^{\star})$. Assumption 2 guarantees the necessary information exchange between agents.

As mentioned in Section \ref{sec:Introduction}, noise caused by transmission of the information may lead to deterioration of system performance or even influence its convergence property which can be reflected in the experimental results in Section \ref{sec:Numerical Experiments}. In this paper, we focus on the error caused by noisy communication links and/or quantization. In the following, two independent random sequences $\{\Xi_k \}$, $\{E_k \}$ are used to summarize the above mentioned noise:
$$\Xi_k :=[\boldsymbol{\xi}_{1,k}, \boldsymbol{\xi}_{2,k},\cdots, \boldsymbol{\xi}_{n,k}]^{\intercal }, \ E_k :=[\boldsymbol{\varepsilon}_{1,k}, \boldsymbol{\varepsilon}_{2,k},\cdots, \boldsymbol{\varepsilon}_{n,k}]^{\intercal }$$
where $\boldsymbol{\xi}_{i,k}$, $\boldsymbol{\varepsilon}_{i,k} \in \mathbb{R}^p$ denote the noise encountered by agent $i$ at iteration time $k$.

In practice, the communication channels can be corrupted by additive noise and was statistically modeled as Gaussian in \cite{Gaussian_noise}. Without loss of generality, in this paper, the noise caused by noisy communication links is considered to be zero mean white Gaussian. To save communication resources, we apply the following random quantization scheme \cite{DSGRQ} to quantize our data. For example, for single number $x \in [l,u]$, we uniformly divide the interval into $B$ bins, whose end points are bounded by $\tau_{i}$, i.e., $l=\tau_{1}\leq\tau_{2}\leq\cdots\leq\tau_{B}= u$ and $\triangle\triangleq\tau_{i+1}-\tau_{i}=\frac{u-l}{B-1}$. Thus, $b=log_2 \left(B\right)$ bits can be used to index the $\{\tau_{i}\}$. Given $x\in \left[\tau_{i},\tau_{i+1}\right) $, we assign a probability based on its relative location inside this interval and choose either $\tau_{i}$ or $\tau_{i+1}$ to represent $x$ at random:
\begin{equation*}
Q(x) = \left\lbrace \begin{array}{ll}\tau_{i} & \text{w.p. } 1-\frac{x-\tau_{i}}{\triangle}, \\
\tau_{i+1} & \text{w.p. } \frac{x-\tau_{i}}{\triangle}. \end{array} \right.
\end{equation*}
where the random variable $Q\left(x\right)$ satisfies
\begin{align*}
\mathbb{E}[Q\left(x\right)]=x, \ \  \mathbb{E}[\left(Q\left(x\right)-x\right)^2]\leq\frac{\triangle^2}{4}
\end{align*}

Therefore, $\Xi_k$, $E_k$ have zero mean and bounded variance, i.e., $\mathbb{E}[E_k]=\mathbb{E}[\Xi_k]=0$ and there exist $\sigma_\varepsilon$, $\sigma_\xi>0$ such that $\mathbb{E}[||E_k||_F^2]\leq \sigma_\varepsilon^2$, $\mathbb{E}[||\Xi_k||_F^2]\leq \sigma_\xi^2$.

Our goal in this paper is to solve the EDP in a distributed manner with noisy information exchange. 

\section{Noise-robust Distributed Dual Gradient Tracking Algorithm}
\label{sec:Algorithm}
In this section, we propose a noise-robust variant of gradient tracking based algorithm to solve the dual problem and consequently the EDP in a fully distributed manner. 

Introducing Lagrange multiplier $\boldsymbol{\lambda }$, we construct its Lagrangian $L\left(W,\boldsymbol{\lambda }\right)$. Then, the dual counterpart of EDP can be derived and it is equivalent to the following in the form of a distributed consensus problem (e.g., see \cite{PUSHPULL5} for details)
\begin{align}
\label{dual_costfunction}
&\underset{\boldsymbol{\lambda }\in \mathbb{R}^p}{\mathop{\text{minimize}}} \quad \sum_{i=1}^n{F_i\left( \boldsymbol{\lambda } \right)}=\sum_{i=1}^n{F_i\left( \boldsymbol{\lambda }_i \right)}\\
&\text{subject to} \quad \boldsymbol{\lambda }_1=\boldsymbol{\lambda }_2=\cdots=\boldsymbol{\lambda }_n=\boldsymbol{\lambda } \in \mathbb{R}^p
\end{align}
in which the local cost function is as follows 
$$
F_i\left(\boldsymbol{\lambda} \right) \triangleq  f_{i}^{*}\left( -\boldsymbol{\lambda} \right) +\boldsymbol{\lambda }^{\intercal } \mathbf{d}_{i}. 
$$
where $f_{i}^{*}\left( \boldsymbol{\lambda } \right)$ is the convex conjugate function corresponding to the pair $\left(f_i,\mathcal{W}_i\right)$ which is given by   
\begin{align}
f_{i}^{*}\left( \boldsymbol{\lambda } \right) =\underset{\mathbf{w}\in \mathcal{W}_i}{\sup}\left( \mathbf{w}^{\intercal } \boldsymbol{\lambda }-f_i\left( \mathbf{w} \right) \right). 
\end{align} 

\emph{Assumption 3:} The problem satisfies Slater’s condition, i.e., there exists at least one point in the relative interior of $\mathcal{W}$ can satisfy the power balance constraint (\ref{power balance constraint}).

Under Assumptions 1, 3, the strong duality holds. Therefore, for the dual problem, its optimal value $F^{\star }=-f^{\star }$.
\subsection{Effects of Noisy Information}
To demonstrate the adverse effects of the noise to algorithms with the gradient tracking scheme, a typical algorithm \cite{firstorder} is shown follows (with noise):
\begin{align}
X_{k+1}&=\left(R X_k+\Xi _k\right)-\alpha Y_k\\
\label{GT}
Y_{k+1}&=\left(CY_k+E _k\right)+\nabla F\left( X_{k+1} \right) -\nabla F\left( X_k \right) 
\end{align}
where initialization should satisfy $Y_{0}=\nabla F\left( X_{0} \right)$ and the matrix $R$ is  nonnegative row-stochastic, $C$ is nonnegative column-stochastic, i.e., $R\mathbf{1}=\mathbf{1}$ and $\mathbf{1}^{\intercal }C=\mathbf{1}^{\intercal }$. $Y_k$ plays the role of tracking the average gradient using 
the column-stochastic matrix $C$. In the case of noiseless communication, it can be derived by left multiplying both sides of (\ref{GT}) by $\frac{1}{n}\mathbf{1}^{\intercal }$ and accumulating the formula from $0$ to $k$ that
\begin{align}
\label{GT_relation}
\frac{1}{n}\mathbf {1}^{\intercal } Y_k=\frac{1}{n}\mathbf {1}^{\intercal }\nabla F(X_{k}) \qquad \forall k.
\end{align}
(\ref{GT_relation}) is the fundamental relation for achieving the so-called gradient tracking through $Y$-update which has been shown in literature (see e.g., \cite{PUSHPULL1}, \cite{PUSHPULL2}). However, noisy information makes (\ref{GT_relation}) no longer hold and leads to a poor performance of the algorithm as we have by induction that   
\begin{align}
\frac{1}{n} \mathbf{1}^{\intercal }Y_k=\frac{1}{n}\mathbf{1}^{\intercal }\left( \nabla F\left( X_k \right) +\sum_{t=0}^{k-1}{E_t} \right) .
\end{align}
where the variance of the summation term of noise tends to infinity as $k$ grows which means instability or low reliability in some way. Hence, there are some works need to be done to eliminate the undesirable summation term of noise. 
\subsection{Development of a Noise-robust Algorithm}
Inspired by the induction process of relation (\ref{GT_relation}) without noise interference, we introduce an auxiliary variable $S_{k+1}$ to record noise's impact on gradient information at iteration $k$ which is named noise-tracing:
\begin{align}
S_{k+1}=CS_k+\nabla F\left( X_{k+1} \right)+E_k
\end{align}
And $\left(S_{k+1}-S_{k}\right)$ is used to the process of tracking average (stochastic) gradient where the effect of the noise at iteration $k-1$ can be eliminated during the iteration $k$. If the similar initialization condition $S_{0}-S_{-1}=\widetilde{\nabla} F\left( X_{0} \right)$ holds, we have
\begin{align}
\frac{1}{n}\mathbf {1}^{\intercal } \left(S_{k+1}-S_{k}\right)=\frac{1}{n}\mathbf {1}^{\intercal } \widetilde{\nabla} F(X_{k+1})
\end{align}
where $\widetilde{\nabla} F(X_{k+1})=\nabla F(X_{k+1})+E_k$ denotes the unbiased estimate of $\nabla F(X_{k+1})$ with bounded variance. Thus, the introduction of $S_{k+1}$ effectively avoids the occurrence of the summation term of noise and transforms the algorithm to a stochastic gradient version. Under Assumption 1, function $f_{i}^{*}\left( \boldsymbol{\lambda } \right)$ is differentiable with Lipschitz continuous gradients \cite{Covex_Optimization}, so is $F_i\left(\boldsymbol{\lambda }\right)$. From the proposition in \cite{Danskins_theorem} we have 
\begin{align}
\nabla F_i(\boldsymbol{\lambda})&=-\nabla f_{i}^{*}\left( \boldsymbol{\lambda } \right)+\mathbf{d}_i \notag \\
\label{dual_gradient}
&=-\underset{\mathbf{w}\in \mathcal{W}_i}{\text{argmin}}\left\{ f_i\left( \mathbf{w} \right) +\mathbf{w}^{\intercal}\boldsymbol{\lambda } \right\}+\mathbf{d}_i.
\end{align}
Some stochastic gradient based algorithms developed recently are shown to work well in literature (see e.g., \cite{sGT}, \cite{sGT1}). So, based on the noise-tracing scheme and relation (\ref{dual_gradient}), a noise-robust distributed dual gradient tracking algorithm (RDDGT) is proposed which is shown in Algorithm 1 in details.

\begin{table}[!ht]
	\renewcommand{\arraystretch}{1.3}
	\centering
	\label{table_example}
	\begin{tabular}{p{8.5cm}}
		\hline
		\textbf{Algorithm 1:} RDDGT --- from the view of agent $i$\\
		\hline
		1:\ Initialization: design $R=[r_{ij}]$, $C=[c_{ij}]$, select proper $\alpha>0$ and $\eta ,\gamma \in \left( 0,1 \right]$ , let $\mathbf{x}_{i,0}=\mathbf{w}_{i,0}=0,$ $\mathbf{s}_{i,0}=\mathbf{d}_i.$\\ 
		2:\ $
		\mathbf{For}\ k=0,1,…,K\ \mathbf{do}
		$\\
		3:\ Dual Update:\\$
		\qquad \qquad \mathbf{x}_{i,k+1}=\left( 1-\eta \right) \mathbf{x}_{i,k}+\eta \left( \sum _{j=1}^n{r_{ij}\cdot \mathbf{x}_{j,k}}+\boldsymbol{\xi} _{i,k} \right) 
		$\\
		$
		\qquad \qquad \qquad \quad \quad \ \ -\alpha \cdot \left( \mathbf{s}_{i,k}-\mathbf{s}_{i,k-1} \right),
		$\\
		4:\ Primal Update:\\ $
		\qquad \qquad \mathbf{w}_{i,k+1}=\mathop{\arg\min}\limits_{\mathbf{w}\in \mathcal{W}_i}\left\{ f_i\left( \mathbf{w} \right) +\mathbf{w}^T\cdot \mathbf{x}_{i,k+1} \right\},
		$\\
		5:\ Auxiliary Update:\\ $
		\qquad \qquad \mathbf{s}_{i,k+1}=\left( 1-\gamma \right) \mathbf{s}_{i,k}+\gamma \left( \sum _{j=1}^n{c_{ij}\cdot \mathbf{s}_{j,k}}+\boldsymbol{\varepsilon}_{i,k} \right) 
		$\\
		$
		\qquad \qquad \qquad \quad \quad  \ 
		\ -\mathbf{w}_{i,k+1}+\mathbf{d}_{i},
		$\\
		6:\ $
		\mathbf{end\ for.}
		$\\
		\hline
	\end{tabular}
\end{table}
Note that $\mathbf{s}_{i,-1}=\mathbf{0}$ for the first dual update and notations have been changed to keep consistency with the above analysis, i.e., $\mathbf{x}_{i,k}=\boldsymbol{\lambda}_{i,k}$. In Algorithm 1, we use two asymmetrical matrices $R$, $C$ to model the information exchange between agents. The weight matrix $R$ is nonnegative row-stochastic, while $C$ is nonnegative column-stochastic. At each iteration, for $i, j \in \mathcal{N}$, each agent $i$ will “pushes” stochastic (or noisy) gradients information $c_{ji}\mathbf{s}_{i,k}$ to its out-neighbors $j \in \mathcal{N}_i^{\text{out}}$ and “pulls” the (noisy) dual variables $\mathbf{x}_{j,k}$ from its in-neighbors $j \in \mathcal{N}_i^{\text{in}}$, respectively. Each component of $X_k$ performs optimization seeking by average consensus while $\left(S_{k+1}- S_k\right)$ is used to track the stochastic gradient. Note that a local optimization problem should be solved per primal update which  is common in many duality-based algorithms such as DuSPA algorithm in \cite{DuSPA}. The subproblem is in fact a constrained minimization problem with strongly convex cost function which can be solved through standard algorithms, e.g., projected gradient methods for general constraints. In some special cases this update can be carried out through specific methods, for example, when $\mathcal{W}_i$ encodes the generation capacity of agent $i$, i.e., $\mathcal{W}_i=[\underline{w}_i,\overline{w}_i]$,
\begin{align}
w_{i,k+1}=\min \left\{ \max \left\{ \nabla ^{-1}f_i\left( -x_{i,k+1} \right) ,\underline{w}_i \right\} ,\overline{w}_i \right\}. 
\end{align}
where $\nabla ^{-1}f_i$ denotes the inverse function of $\nabla f_i$ and $\underline{w}_i ,\overline{w}_i$ denote the lower and upper limits of local power generation, respectively.

In addition to the stepsize $\alpha$ for gradient descent, we introduce parameters $\eta$, $\gamma$ that determine the degree to which variables from the neighbors should be weighed against the local one when proceeding algorithm updating. That is, using local information to reduce the impact of the random noise. Or they can be simply seen as averaging parameters between weight matrices and identity matrix $\mathbf{I}$.

Let $Y_{k+1}=S_{k+1}-S_k$, $R_\eta=[r_{\eta,ij}]=\left(1-\eta\right) \mathbf{I}+\eta R$, $C_\gamma=[c_{\gamma,ij}]=\left(1-\gamma\right) \mathbf{I}+\gamma C$. The proposed algorithm can be written as follows
\begin{align}
&\mathbf{x}_{i,k+1}=\sum _{j=1}^n{r_{\eta,ij}\cdot \mathbf{x}_{j,k}}+\eta  \boldsymbol{\xi} _{i,k} -\alpha \mathbf{y}_{i,k}\\
&\mathbf{w}_{i,k+1}=\mathop{\arg\min}\limits_{\mathbf{w}\in \mathcal{W}_i}\left\{ f_i\left( \mathbf{w} \right) +\mathbf{w}^T\cdot \mathbf{x}_{i,k+1} \right\}\\
&\mathbf{y}_{i,k+1}=\sum _{j=1}^n{c_{\gamma,ij}\cdot \mathbf{y}_{j,k}}+[\widetilde{\nabla} F_i(\mathbf{x}_{i,k+1})-\widetilde{\nabla} F_i(\mathbf{x}_{i,k})]
\end{align}
where $\widetilde{\nabla} F_i(\mathbf{x}_{i,k+1})={\nabla} F_i\left(\mathbf{x}_{i,k+1}\right)+\gamma \boldsymbol{\varepsilon}_{i,k}$. In a distributed manner, the algorithm is carried out in the way that: all the variables $\mathbf{x}_{i,k}$ are driven to a consensus value $\overline{\mathbf{x}}_k$ while $\overline{\mathbf{x}}_k$ is driven to the optimum $\mathbf{x}^\star$ at the same time. Note that despite the strongly convexity of $f_i\left(\mathbf{w}_i\right)$, the cost function $F_i\left(\boldsymbol{\lambda }\right)$ in (\ref{dual_costfunction}) is often not strongly convex due to the introduction of conjugate function $f_i^{*}\left(\boldsymbol{\lambda }\right)$, e.g., $f_i$ includes exponential term. Inspired by existing works \cite{sGT1}, \cite{RPPG}, its convergence property is quantified by the following term:
\begin{align*}
M\left(k\right)=\frac{1}{k} \sum\limits_{t=1}^k{a\cdot \mathbb{E}\left[ ||\nabla F\left( \mathbf{\bar{x}}_t \right) ||^2 \right] +b\cdot \mathbb{E}\left[ ||X_t-1\mathbf{\bar{x}}_{t}^{T}||_F^2 \right]}
\end{align*}
where $\mathbb{E}\left[ ||\nabla F\left( \mathbf{\bar{x}}_t \right) ||^2 \right]$ measures the optimality gap to the optimum and $\mathbb{E}\left[ ||X_t-1\mathbf{\bar{x}}_{t}^{T}||_F^2 \right]$ measures the consensus error among agents. Constant parameters $a$, $b \in \mathbb{R}^+$ are arbitrary bounded positive real numbers. The result is summarized in Theorem 1 whose proof is omitted due to space limitations.

\emph{Theorem 1:} Suppose Assumptions 1-3 hold and the cost function $F_i$, $i \in \mathcal{N}$ is convex, differentiable and $L$-Lipschitz gradient continuous (c.f. (\ref{dual_gradient})). With proper chosen $\eta $, $\gamma \in \left(0, 1\right]$, if the stepsize $\alpha$ is sufficiently small, the variable $\{\mathbf{x}_{i,k}\}$ generated by RDDGT satisfies
\begin{align}
\label{theorem1}
M\left(k\right) \leq C_1\frac{1}{k}+C_2 \sigma_\varepsilon^2+C_3 \sigma_\xi^2
\end{align}
where parameters $C_1$, $C_2$, $C_3$ are independent of iteration step $k$ which are determined by weight matrices as well as the selection of suppression parameters and the stepsize. We bound $M\left(k\right)$ by a linear programming (LP) based method. Three complicated but important terms in the deduction process are bounded by three interlacing linear inequalities which are regarded as constraints of LP and their positive linear combination as the objective function. By explicitly solving the LP problem, an upper bound of $M\left(k\right)$ can be derived.

From (\ref{theorem1}), we obtain that $F\left(X_k\right)=\sum_{i=1}^n{F_i\left( \mathbf{x}_i \right)}$ converges to the neighborhood of optimal value $F^\star$ by the convexity, i.e., there exist a minimum bounded constant $\varGamma >0$ satisfies $F\left( X_k \right) -F^{\star}\le \varGamma$ when iteration step $k$ is sufficiently large. 

\emph{Theorem 2:} Suppose Assumptions 1-3 hold and the stepsize $\alpha$ is smaller than the upper bound given in \emph{Theorem 1}. The variable $\{\mathbf{w}_{i,k}\}$ generated by RDDGT satisfies 
\begin{align}
\label{theorem2}
\Vert W^{\star}-W_k\Vert^2\le \frac{2}{\mu}\varGamma. 
\end{align}

\emph{Proof:} Under Assumptions 3, the strong duality holds and thus the optimal point $W^{\star}$ and the optimal value of the primal problem $f^{\star}$ satisfy that $F^{\star}=-f^{\star}=-L\left( W^{\star},\mathbf{x} \right) , \forall \mathbf{x}\in \mathbb{R}^p$.

At iteration $k$, we have
\begin{align}
F\left( \mathbf{\bar{x}}_k \right) -F^{\star}=&\left( -\underset{W\in \mathcal{W}}{inf}L\left( W,\mathbf{\bar{x}}_k \right) \right) -\left( -L\left( W^{\star},\mathbf{\bar{x}}_k \right) \right) \notag \\
=&L\left( W^{\star},\mathbf{\bar{x}}_k \right) -L\left( W_k,\mathbf{\bar{x}}_k \right)
\label{theorem2_1}
\end{align}
 where $W_k$ is the optimal point of $\underset{W\in \mathcal{W}}{inf}L\left( W,\mathbf{\bar{x}}_k \right)$. So, we have 
 \begin{align}
 -\partial _WL\left( W_k,\mathbf{\bar{x}}_k \right) ^T\left( W^{\star}-W_k \right) \leq 0.
 \label{theorem2_2}
 \end{align}
 
 Moreover, due to the strongly convexity of $f$, the Lagrangian $L\left(W,\boldsymbol{\lambda }\right)$ is also strongly convexity:
 \begin{align}
 L\left( W^{\star},\mathbf{\bar{x}}_k \right) \geq& L\left( W_k,\mathbf{\bar{x}}_k \right) +\partial _WL\left( W_k,\mathbf{\bar{x}}_k \right) ^{\intercal}\left( W^{\star}-W_k \right) \notag \\
 &+\frac{\mu}{2}\Vert W^{\star}-W_k\Vert ^2.
 \label{theorem2_3}
 \end{align}
When iteration time $k$ is sufficiently large, it follows from $\left(\ref{theorem2_1}\right)$, $\left(\ref{theorem2_2}\right)$, $\left(\ref{theorem2_3}\right)$ that
\begin{align}
\Vert W^{\star}-W_k\Vert ^2\leq & \frac{2}{\mu}\left( L\left( W^{\star},\mathbf{\bar{x}}_k \right) -L\left( W_k,\mathbf{\bar{x}}_k \right) \right) \notag  \\
=& \frac{2}{\mu}\left( F\left( \mathbf{\bar{x}}_k \right) -F^{\star} \right) \le \frac{2}{\mu}\varGamma 
\end{align}

In a special case when $\mathcal{W}=\mathbb{R}^m$ and cost functions of every nodes $f_i$ is differentiable, the following theorem gives the convergence rate result for our proposed algorithm. Under Assumtions 1 and 3, we have the necessary and sufficient condition for optimality from the KKT condition:
\begin{align}
&\nabla f_1\left( \mathbf{w}_{1}^{\star} \right) =\nabla f_2\left( \mathbf{w}_{2}^{\star} \right) =...=\nabla f_n\left( \mathbf{w}_{n}^{\star} \right) \label{KKT_1} \\
&\sum_{i=1}^n{\mathbf{w}_{i}^{\star}}=\sum_{i=1}^n{\mathbf{d}_i}=D \label{KKT_2}
\end{align}

Thus, we choose the following term to make analysis
\begin{align*}
N\left(k\right)=\mathbb{E}\left[ \sum_{i=1}^n{\Vert \nabla f_i\left( \mathbf{w}_{i,k} \right) -\frac{1}{n}\sum_{j=1}^n{\nabla f_j\left( \mathbf{w}_{j,k} \right)}\Vert ^2} \right] \\ +\mathbb{E}\left[\Vert \sum_{i=1}^n{\mathbf{w}_{i,k}}-D||^2 \right] 
\end{align*}  
where two part correspond to $\left(\ref{KKT_1}\right)$ and $\left(\ref{KKT_2}\right)$, respectively.

\emph{Theorem 3:} Suppose Assumptions 1-3 hold and the stepsize $\alpha$ is smaller than the upper bound given in \emph{Theorem 1}. If $\mathcal{W}=\mathbb{R}^m$ and cost functions $f_i$ are all differentiable. The variable $\{\mathbf{w}_{i,k}\}$ generated by RDDGT satisfies 
\begin{align}
\label{theorem3}
N\left(k\right) \leq D_1\frac{1}{k}+D_2 \sigma_\varepsilon^2+D_3 \sigma_\xi^2
\end{align}
where parameters $D_1$, $D_2$, $D_3$ are independent of iteration step $k$. The proof of \emph{Theorem 3} is deferred to the Appendix.

Thus, on the basis of Theorem 1, using the strongly convexity of $f_i$, we can derived that decision variables $\mathbf{w}_i$ can also converge to a neighborhood of the optimum. Thus, the convergence of the proposed algorithm is established. And, it is proved that the algorithm has a sublinear convergence rate $O\left(\frac{1}{k}\right)$ which matches existing algorithms. Our proposed algorithm drives the solution to converge to the optimum's  neighborhood which is directly related to the variance $\sigma_\varepsilon$, $\sigma_\xi$ and the neighborhood can be tighter through proper tuning while other algorithms may diverge.
\subsection{Algorithm Acceleration}
\label{sec:Algorithm Improvement}
In this section, we develop an accelerated variant of RDDGT by introducing a momentum. Nesterov-type acceleration was first proposed in \cite{Nesterov} by Nesterov which was originally aimed at accelerating gradient descent-type (first-order) methods. Recently, this technique has been extended to wider applications, such as ADMM-based algorithms \cite{ADMMN}, gradient tracking-based algorithms \cite{ABN}. Inspired by these works, the acceleration version of RDDGT with Nesterov momentum named RDDGT-N can be derived where $\beta$ is the momentum parameter and the $\mathbf{x}$-update is changed to a two-step process:
\begin{align}
&\hat{\mathbf{x}}_{i,k+1}=\sum _{j=1}^n{r_{\eta,ij}\cdot \mathbf{x}_{j,k}}+\eta  \boldsymbol{\xi} _{i,k} -\alpha \mathbf{y}_{i,k}\\
&\mathbf{x}_{i,k+1}=\hat{\mathbf{x}}_{i,k+1}+\beta \left(\hat{\mathbf{x}}_{i,k+1}-\hat{\mathbf{x}}_{i,k}\right)
\end{align}

The effectiveness of the aforementioned strategy is validated through numerical experiments in Section \ref{sec:Numerical Experiments}.

\section{Numerical Experiments}
\label{sec:Numerical Experiments}
In this section, we present numerical experiments to examine the effectiveness of RDDGT and its accelerated version in the IEEE 14-bus system. RDDGT is also applied to the IEEE 57-bus power system to demonstrate its scalability. The optimization has been run on a computer with Intel-i5 1.8 GHz CPU and 8 GB of RAM. We consider an EDP with generation capacity limitation as its local constraint as mentioned in Section \ref{sec:Algorithm}, i.e., $\mathcal{W}_i=[\underline{w}_i, \overline{w}_i]$. Note that if a bus has no generator, we can set $\underline{w}_i= \overline{w}_i=0$ for simplicity. As in \cite{cost_function}, the cost function is modeled as a quadratic function:
\begin{align*}
f_i\left(w_i\right)=a_i\left(w_i\right)^2+b_i w_i+c_i
\end{align*} 

The parameters adopted from \cite{DuSPA} can be found in Table \ref{parameters_table}.

\begin{table}[!ht]
	\renewcommand{\arraystretch}{1.3}
	\centering
	\caption{Generator Parameters}
	\label{parameters_table}
	\begin{tabular}{|c|c|c|c|c|}
		\hline
		BUS & a$\left(\$ / MW^2 \right)$ & b$\left(\$ / MW \right)$ & c$\left(\$\right)$ & $[\underline{w}_i, \overline{w}_i]$ $\left(MW\right)$ \\
		\hline
		1 & 0.04 & 2.0 & 0 & $[0,80]$\\
		\hline
		2 & 0.03 & 3.0 & 0 & $[0,90]$\\
		\hline
		3 & 0.035 & 4.0 & 0 & $[0,70]$\\
		\hline
		6 & 0.03 & 4.0 & 0 & $[0,70]$\\
		\hline
		8 & 0.04 & 2.5 & 0 & $[0,80]$\\
		\hline
	\end{tabular}
\end{table}

To design proper weight matrices, for any agent $i \in \mathcal{N}$, we use $|\mathcal{N}_i^{in}|$, $|\mathcal{N}_i^{out}|$ to denote the number of its in-neighbors and out-neighbors, respectively. The weight matrices $R$ and $C$ can be obtained as follows: let $r_{i,j}= \frac{1}{|\mathcal{N}_i^{in}|+1}$ for all $j \in \mathcal{N}_i^{in}$ and thus $r_{i,i}= 1-\sum_{j \in \mathcal{N}_i^{in}}r_{i,j}$. Similarly, $c_{j,i}= \frac{1}{|\mathcal{N}_i^{out}|+1}$ for all $j \in \mathcal{N}_i^{out}$ and thus $c_{i,i}= 1-\sum_{j \in \mathcal{N}_i^{out}} c_{j,i}$.
\subsection{Convergence of the Proposed Algorithm}
\begin{figure}[!ht]
	\centering
	\subfloat[]{\includegraphics[width=3in]{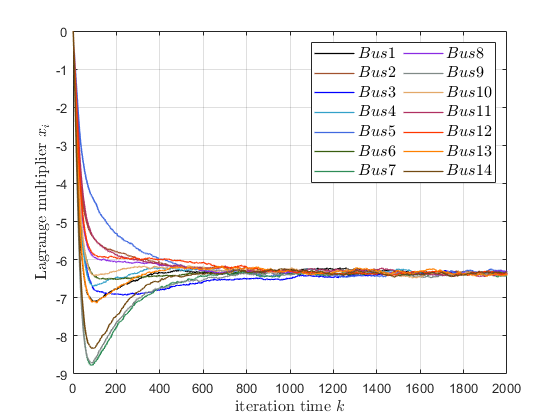}
		\label{OPT_fig_Con}}\\
	\subfloat[]{\includegraphics[width=3in]{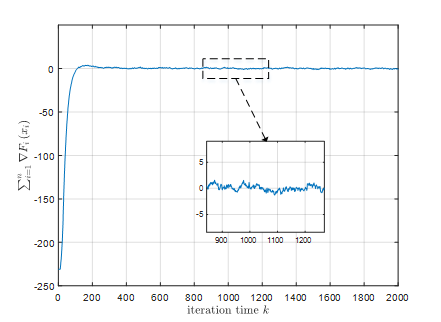}
		\label{OPT_fig_Gra}}
	\caption{The historical evolution of $x_i$ and $\sum_{i=1}^n{\nabla F_i\left( x_i \right)}$ with respect to iteration time $k$.}
	\label{OPT_Agent}
\end{figure}

\begin{figure}[!ht]
	\centering
	\includegraphics[width=3in]{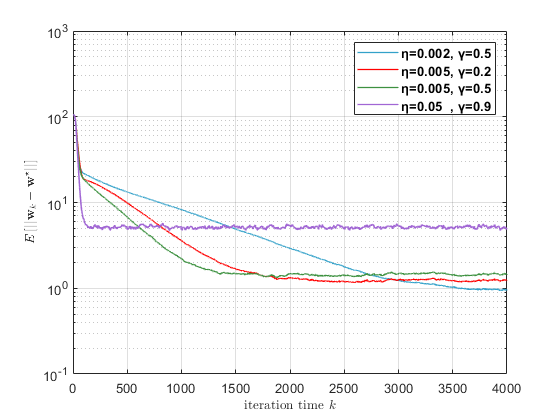}
	\caption{ Convergence property under different suppression parameters setting.}
	\label{Diffpar_fig}
\end{figure}

Set $\alpha = 0.01$, total load demand $d_{total}=231MW$ , the information exchange process is assumed be corrupted with independent Gaussian noise $\mathcal{N}\left(0,1\right)$ and we apply RDDGT to EDP. Fig.~\ref{OPT_Agent}(a) and \ref{OPT_Agent}(b) plot the historical evolution of the Lagrange multiplier $x_i$ and gradient information $\sum_{i=1}^n{\nabla F_i\left( x_i \right)}$ in order to demonstrate the consensus and optimality of the proposed algorithm, respectively. The results are shown to be consistent with Theorem 1 in Section \ref{sec:Algorithm}. Next, we examine the effect of the suppression parameters $\eta$, $\gamma$ on the convergence property of RDDGT. Fig.~\ref{Diffpar_fig} shows how the error $\mathbb{E}\left[ ||\mathbf{w}_k-\mathbf{w^{\star}}||  \right] $ changes with iteration time $k$, where the expected errors are approximated by averaging over 100 simulation results. It follows from the figure that large suppression parameters can lead to a lower convergence time but a higher error.

\subsection{Algorithm Acceleration}
\begin{figure}[!ht]
	\centering
	\includegraphics[width=3in]{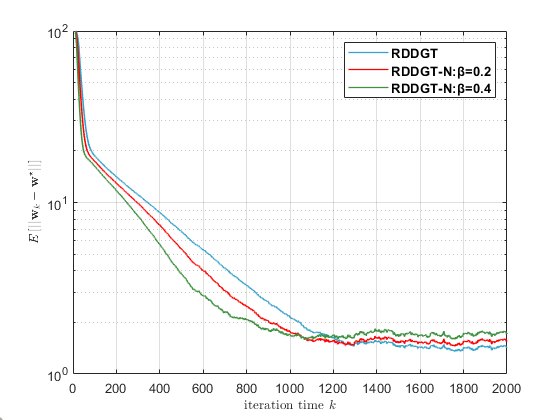}
	\caption{ Performance comparison of the proposed algorithm with and without the proposed acceleration strategy.}
	\label{SPD_fig}
\end{figure}

Then, we examine the effectiveness of the additional momentum term on accelerating our algorithm. We choose momentum parameters $\beta=0.2, \ 0.4$, Fig.~\ref{SPD_fig} depicts the simulation result of the error with iteration time $k$. It follows from the figure that the acceleration strategy can lead to a faster convergence and only has acceptable effects on the error. 
\begin{figure}[!ht]
	\centering
	\subfloat[]{\includegraphics[width=3in]{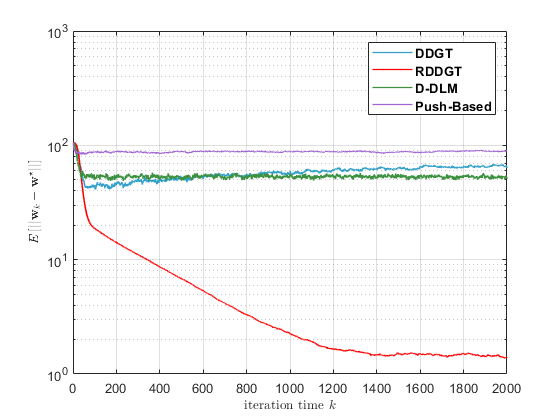}
		\label{OPT_Diffalg}}\\
	\subfloat[]{\includegraphics[width=3in]{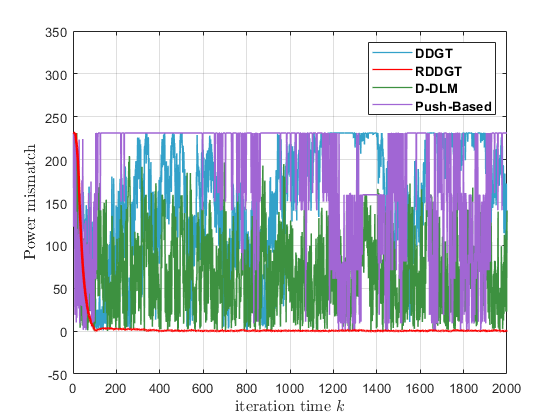}
		\label{OPT_Diffalg_PM}}
	\caption{The historical evolution of $\mathbb{E}\left[ ||\mathbf{w}_k-\mathbf{w^{\star}}||  \right] $ and total power mismatch with respect to iteration time $k$.}
	\label{OPT_fig}
\end{figure}
\subsection{Comparison with the State-of-the-art}
We compare the proposed RDDGT with three state-of-the-art algorithms, termed Push-Based algorithm \cite{Push_Based}, D-DLM \cite{D_DLM} and DDGT \cite{PUSHPULL5}, in terms of the error $\mathbb{E}\left[ ||\mathbf{w}_k-\mathbf{w^{\star}}||  \right] $ which is shown in Fig.~\ref{OPT_fig}. It can be seen that initially all the errors decrease at comparable rates. However, the errors of competitors tend to increase while our proposed algorithm can converge to a neighborhood of optimum. It follows from the figure that, under noisy information exchange, the power balance constraint in (\ref{power balance constraint}) is achieved by RDDGT fast and accurately while the others can hardly satisfy it. This sharp contrast verifies the effectiveness of the proposed algorithm.

\subsection{Performance of Scalability}
Finally, we examine the scalability of RDDGT in the IEEE 57-bus system. It follows from the figure that the performance of RDDGT is similar to that in IEEE 14-bus system and it still outperforms other algorithms in noise suppression. These plots show the capability of RDDGT to solve the EDP problem even when the network becomes larger. This feature makes it adequate for large power systems problems.
\begin{figure}[!ht]
	\centering
	\subfloat[]{\includegraphics[width=3in]{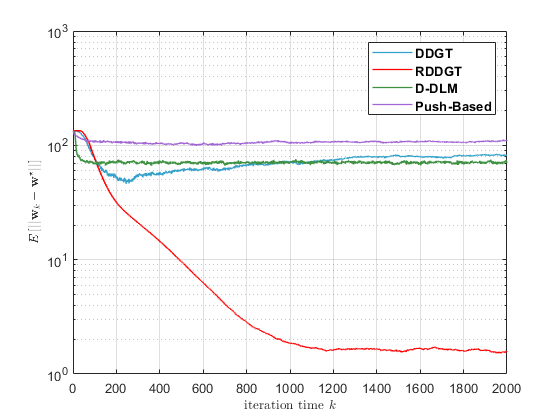}
		\label{OPT57_Diffalg}}\\
	\subfloat[]{\includegraphics[width=3in]{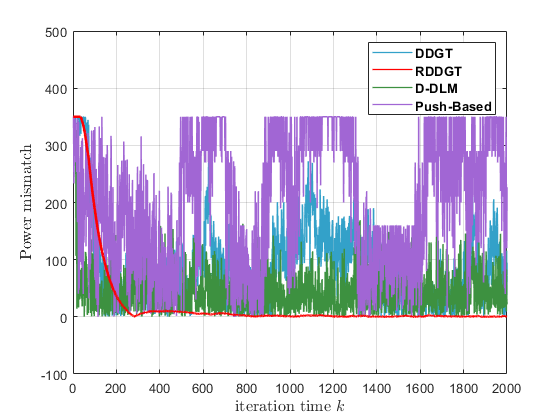}
		\label{OPT57_Diffalg_PM}}
	\caption{The historical evolution of $\mathbb{E}\left[ ||\mathbf{w}_k-\mathbf{w^{\star}}||  \right] $ and total power mismatch with respect to iteration time $k$.}
	\label{OPT57_fig}
\end{figure}

\section{Conclusion}
\label{sec:Conclusion}
In this paper, a noise-robust distributed dual gradient tracking algorithm (RDDGT) has been proposed. And its convergence property for strongly convex cost function has been established under standard assumptions. Under noisy information exchange, RDDGT has been shown to be more robust than existing distributed algorithms for economic dispatch problem. Moreover, an acceleration strategy has been presented for RDDGT by introducing the Nesterov momentum and it is numerically validated. The effectiveness and scalability of the proposed algorithm has been examined by numerical experiments.
\section{Acknowledgment}
The work was supported by the National Natural Science Foundation of China under Grants 61922058, 62173225 and the Key Program of the National Natural Science Foundation of China under Grant 62133008.


%

\bibliographystyle{IEEEtran}
\bibliography{ref}

\section{Appendix}
\subsection{Preliminary results for the proof}
\emph{Definition 1:} Given an arbitary vector norm $\Vert \cdot \Vert$ on $\mathbb{R}^n$, for any $X \in \mathbb{R}^{n\times p} $, we define
\begin{align*}
 \Vert X \Vert = \Vert \left[ \Vert X^{\left(1\right)} \Vert , \Vert X^{\left(2\right)}\Vert ,\dots ,\Vert X^{\left(p\right)}\Vert \right] \Vert _2
\end{align*}

The following four lemmas are taken from \cite{PUSHPULL}.

\emph{Lemma 1:} Under Assumption 2 and suppose matrix $R$, $C$ satisfy $R\mathbf{1}=\mathbf{1}$, $\mathbf{1}^{\intercal}C=\mathbf{1}^{\intercal}$, matrix $R$ has unique nonnegative left eigenvector $\mathbf{u}^{\intercal}$ w.r.t. eigenvalue 1 with $\mathbf{u}^{\intercal} \mathbf{1}=n$; matrix $C$ has unique nonnegative right eigenvector $\mathbf{v}$ w.r.t. eigenvalue 1 with $\mathbf{1}^{\intercal}\mathbf{v}=n$ and $\mathbf{u}^{\intercal}\mathbf{v}>0$.

\emph{Lemma 2:} Under Assumption 2 and suppose matrix $R$, $C$ satisfy $R\mathbf{1}=\mathbf{1}$, $\mathbf{1}^{\intercal}C=\mathbf{1}^{\intercal}$, there exsit matrix norms $\Vert \cdot\Vert _R$ and $\Vert \cdot\Vert _C$ such that
\begin{align*}
\tau_R:=\Vert R_{\eta}-\frac{\mathbf{1}\mathbf{u}^{\intercal}}{n}\Vert _R<1, \tau_C:=\Vert C_{\gamma}-\frac{\mathbf{v}\mathbf{1}^{\intercal}}{n}\Vert _C<1.
\end{align*}
From lemma 1,2, we know that if matrix $R$, $C$ is determined, $\tau_R$, $\tau_C$ are only related to $\eta$, $\gamma$, respectively. And $\mathbf{u}^{\intercal} R_{\eta}=\mathbf{u}^{\intercal}$, $ C_{\gamma}\mathbf{v}=\mathbf{v}$ hold.

\emph{Lemma 3:} Given an arbitrary norm $\Vert \cdot \Vert$, for any matrix $W \in \mathbb{R}^{n \times n}$ and vector $\mathbf{x} \in \mathbb{R}^{n \times p}$, we have $\Vert W \mathbf{x} \Vert \leq \Vert W\Vert \Vert \mathbf{x} \Vert$. For any $w \in \mathbb{R}^{n \times 1}$ and $x \in \mathbb{R}^{1 \times p} $, we have $\Vert wx\Vert =\Vert w\Vert \Vert x\Vert _2$.

\emph{Lemma 4:} There exist constants $\delta _{R,2}$, $\delta _{C,2}$ such that for any matrix $X \in \mathbb{R}^{n \times p}$, we have $\Vert X \Vert _2 \leq \Vert X \Vert_R$, $\Vert X \Vert _R \leq \delta _{R,2} \Vert X \Vert_2$, $\Vert X \Vert _2 \leq \Vert X \Vert_C$, $\Vert X \Vert _C \leq \delta _{C,2} \Vert X \Vert_2$.
\subsection{Proof of Theorem 1:}
We first define the following variables:
\begin{align}
\bar{\mathbf{x}}_k:=\frac{1}{n}\mathbf{u}^{\intercal}\mathbf{x}, \ \ ,\bar{\mathbf{y}}_k:=\frac{1}{n}\mathbf{1}^{\intercal}\mathbf{y}_k.
\label{definition_bar}
\end{align}
It follows $\left(\ref{definition_bar}\right)$ that
\begin{align*}
\bar{\mathbf{x}}_{k+1}-\bar{\mathbf{x}}_{k} &=\frac{1}{n}\left[\left(R_{\eta} X_k - I \right) + \eta \Xi_k -\alpha Y_k\right]^{\intercal} \\
&= - \frac{\alpha}{n}{Y_k}^{\intercal} \mathbf{u}+ \frac{\eta}{n} {\Xi_k}^{\intercal} \mathbf{u}.
\end{align*}
Note that the function $F_i$ is $L$-Lipschitz smooth ($L=\frac{1}{\mu}$) due to the strongly convexity of $f_i$. Thus, we have
\begin{align}
&F\left(\bar{\mathbf{x}}_{k+1}\right) \notag \\
\leq &F\left(\bar{\mathbf{x}}_{k}\right)+ \nabla F\left(\bar{\mathbf{x}}_{k}\right)^{\intercal} \left(\bar{\mathbf{x}}_{k+1}-\bar{\mathbf{x}}_{k}\right)+\frac{L}{2} \Vert \bar{\mathbf{x}}_{k+1}-\bar{\mathbf{x}}_{k} \Vert^2 \notag\\
= &F\left(\bar{\mathbf{x}}_{k}\right)- \frac{\alpha}{n} \nabla F\left(\bar{\mathbf{x}}_{k}\right)^{\intercal} {Y_k}^{\intercal} \mathbf{u}+ \frac{\eta}{n} \nabla F\left(\bar{\mathbf{x}}_{k}\right)^{\intercal} {\Xi_k}^{\intercal} \mathbf{u}  \notag\\
&+ \frac{L \alpha^2}{2n^2} \Vert {Y_k}^{\intercal} \mathbf{u} \Vert ^2+ \frac{L \eta^2}{2n^2} \Vert {\Xi_k}^{\intercal} \mathbf{u} \Vert ^2 +\frac{L \eta \alpha }{n^2} \Vert \mathbf{u} \Vert ^2 Y_k \Xi_k^{\intercal} \notag\\
\leq & F\left(\bar{\mathbf{x}}_{k}\right)- \frac{\alpha}{n} \nabla F\left(\bar{\mathbf{x}}_{k}\right)^{\intercal} {Y_k}^{\intercal} \mathbf{u}+\frac{\eta}{n} \nabla F\left(\bar{\mathbf{x}}_{k}\right)^{\intercal} {\Xi_k}^{\intercal} \mathbf{u}    \notag\\
&+ \frac{L \alpha^2}{n^2} \Vert \mathbf{u} \Vert ^2 \Vert Y_k-\mathbf{v}\bar{\mathbf{y}}_k \Vert^2 +\frac{L \alpha^2}{n^2}\left(\mathbf{v}^{\intercal} \mathbf{u}\right)^2 \Vert \bar{\mathbf{y}}_k \Vert^2 \notag\\
&+\frac{L \eta^2}{2n^2}  \Vert \mathbf{u} \Vert ^2 \Vert {\Xi_k} \Vert^2 +\frac{L \eta \alpha }{n^2} \Vert \mathbf{u} \Vert ^2 Y_k {\Xi_k}^{\intercal}
\label{Theorem1_1}
\end{align}
where the last inequality follows Lemma 3 and the relation $\Vert X + Y \Vert^2 \leq 2\Vert X \Vert^2+2\Vert Y \Vert^2$.

Define $\tilde{\alpha}:=\frac{\alpha}{n}\left(\mathbf{v}^{\intercal} \mathbf{u}\right)$ and $\tilde{\alpha}>0$ holds from Lemma 1. With Lipschitz smoothness, the second part of equation $\left(\ref{Theorem1_1}\right)$ can be bounded as follows
\begin{align}
-& \frac{\alpha}{n} \nabla F\left(\bar{\mathbf{x}}_{k}\right)^{\intercal} {Y_k}^{\intercal} \mathbf{u}\notag\\
=&- \frac{\alpha}{n} \nabla F\left(\bar{\mathbf{x}}_{k}\right)^{\intercal} \left({Y_k-\mathbf{v}\bar{\mathbf{y}}_k}^{\intercal}\right)^{\intercal} \mathbf{u}-\frac{\alpha}{n}\left(\mathbf{v}^{\intercal}\mathbf{u}\right) \Vert \nabla F\left(\bar{\mathbf{x}}_k\right) \Vert^2 \notag\\
&- \frac{\alpha}{n} \left(\mathbf{v} ^{\intercal} \mathbf{u}\right) \nabla F\left(\bar{\mathbf{x}}_{k}\right)^{\intercal}\left(\bar{\mathbf{y}}_k-\nabla F\left(\bar{\mathbf{x}}_k\right)\right) \notag\\
=&-\frac{\tilde{\alpha}}{\mathbf{v}^{\intercal} \mathbf{u}} \nabla F \left(\bar{\mathbf{x}}_k\right)^{\intercal}\left({Y_k-\mathbf{v}\bar{\mathbf{y}}_k}^{\intercal}\right)^{\intercal} \mathbf{u} - \tilde{\alpha} \Vert \nabla F \left(\bar{\mathbf{x}}_k\right) \Vert^2\notag\\
&-\tilde{\alpha} \nabla F\left(\bar{\mathbf{x}}_k\right)^{\intercal}\left(\frac{1}{n}\mathbf{1}^{\intercal} \left(\nabla F\left(X_k\right)-\nabla F\left(\mathbf{1}{\bar{\mathbf{x}}_k}^{\intercal}\right)\right)\right)^{\intercal}  \notag\\
& - \tilde{\alpha} \frac{\gamma}{n}\nabla F\left(\bar{\mathbf{x}}_k\right)^{\intercal}  E_k^{\intercal}\mathbf{1} \notag \\
\leq& \frac{\tilde{\alpha}}{\mathbf{v}^{\intercal} \mathbf{u}} \Vert\nabla F\left(\bar{\mathbf{x}}_k\right)\Vert \cdot \Vert Y_k -\mathbf{v} \bar{\mathbf{y}}_k^{\intercal} \Vert \Vert \mathbf{u} \Vert - \tilde{\alpha} \Vert \nabla F \left(\bar{\mathbf{x}}_k\right) \Vert^2  \notag\\
&+ \frac{L \tilde{\alpha}}{\sqrt{n}} \Vert\nabla F\left(\bar{\mathbf{x}}_k\right)\Vert \cdot \Vert X_k- \mathbf{1} \bar{\mathbf{x}}_k^{\intercal} \Vert- \tilde{\alpha} \frac{\gamma}{n}\nabla F\left(\bar{\mathbf{x}}_k\right)^{\intercal}  E_k^{\intercal}\mathbf{1}  \notag\\
\leq& 2L^2 n \tilde{\alpha} \cdot \Vert X_k- \mathbf{1} \bar{\mathbf{x}}_k^{\intercal} \Vert +\frac{2\tilde{\alpha}}{\left(\mathbf{v}^{\intercal} \mathbf{u}\right)^2} \Vert \mathbf{u} \Vert^2 \cdot \Vert Y_k -\mathbf{v} \bar{\mathbf{y}}_k^{\intercal} \Vert \notag\\
&-\frac{3\tilde{\alpha}}{4} \Vert \nabla F \left(\bar{\mathbf{x}}_k\right) \Vert^2- \tilde{\alpha} \frac{\gamma}{n}\nabla F\left(\bar{\mathbf{x}}_k\right)^{\intercal}  E_k^{\intercal}\mathbf{1}.
\label{Theorem1_2}
\end{align}

Combining $\left(\ref{Theorem1_1}\right)$, $\left(\ref{Theorem1_2}\right)$ and taking exception, we obtain that
\begin{align}
&\mathbb{E}\left[F\left(\bar{\mathbf{x}}_{k+1}\right)\right] \notag \\
\leq& \mathbb{E}\left[F\left(\bar{\mathbf{x}}_{k}\right)\right] +\frac{L \eta^2}{2 n^2} \Vert \mathbf{u} \Vert^2 \sigma_\varepsilon^2 + \frac{11}{4}L^2n \tilde{\alpha} \cdot \mathbb{E}\left[\Vert X_k- \mathbf{1} \bar{\mathbf{x}}_k^{\intercal} \Vert^2\right] \notag \\
&+ \frac{\left(2+L\tilde{\alpha}\right )\tilde{\alpha}}{\left(\mathbf{v}^{\intercal}\mathbf{u}\right)^2}\Vert \mathbf{u}\Vert^2 \cdot \mathbb{E}\left[\Vert Y_k -\mathbf{v} \bar{\mathbf{y}}_k^{\intercal} \Vert\right] +L\tilde{\alpha} \cdot \mathbb{E} \left[\Vert\bar{ \mathbf{y}}_k \Vert^2\right] \notag \\
&-\frac{3 \tilde{\alpha}}{4} \cdot \left(  \mathbb
{E}\left[\Vert\nabla F\left(\bar{\mathbf{x}}_k\right)\Vert\right]+L^2n \tilde{\alpha} \cdot \mathbb{E}\left[\Vert X_k- \mathbf{1} \bar{\mathbf{x}}_k^{\intercal} \Vert^2\right] \right). 
\label{Theorem1_3}
\end{align}

Summing the equation $\left(\ref{Theorem1_3}\right)$ from $1$ to $k$, we can obtain that
\begin{align}
&\sum \limits _{t=1}^{k} \mathbb
{E}\left[\Vert\nabla F\left(\bar{\mathbf{x}}_t\right)\Vert\right]+L^2n \tilde{\alpha} \cdot \mathbb{E}\left[\Vert X_t- \mathbf{1} \bar{\mathbf{x}}_t^{\intercal} \Vert^2\right] \notag \\
\leq & \frac{4\left[F\left(\bar{\mathbf{x}}_1\right)-F^{\star}\right]}{3\tilde{\alpha}}+k\frac{2L \eta^2}{3 n^2 \tilde{\alpha}} \Vert \mathbf{u} \Vert^2 \sigma_\varepsilon^2 \notag \\
&+\frac{4}{3}L \tilde{\alpha}\cdot \sum \limits _{t=1}^{k} \mathbb
{E}\left[\Vert \bar{\mathbf{y}}_t \Vert ^2\right] + \frac{11}{3}L^2n \cdot \sum \limits _{t=1}^{k} \mathbb
{E}\left[\Vert X_t- \mathbf{1} \bar{\mathbf{x}}_t^{\intercal} \Vert^2\right] \notag\\
&+ \frac{4\left(2+L\tilde{\alpha}\right )}{3\left(\mathbf{v}^{\intercal}\mathbf{u}\right)^2}\Vert \mathbf{u}\Vert^2 \cdot \sum \limits _{t=1}^{k} \mathbb
{E}\left[\Vert Y_t -\mathbf{v} \bar{\mathbf{y}}_t^{\intercal} \Vert\right]. 
\label{target_inequality}
\end{align}

From Lemma 4, we define the term $M$ as follows which upper bounded the sum of the last three terms in $\left(\ref{target_inequality}\right)$:
\begin{align}
M=&\frac{4}{3}L \tilde{\alpha}\cdot \sum \limits _{t=1}^{k} \mathbb
{E}\left[\Vert \bar{\mathbf{y}}_t \Vert ^2\right] + \frac{11}{3}L^2n \cdot \sum \limits _{t=1}^{k} \mathbb
{E}\left[\Vert X_t- \mathbf{1} \bar{\mathbf{x}}_t^{\intercal} \Vert_R^2\right] \notag\\
&+\frac{4\left(2+L\tilde{\alpha}\right )}{3\left(\mathbf{v}^{\intercal}\mathbf{u}\right)^2}\Vert \mathbf{u}\Vert^2 \cdot \sum \limits _{t=1}^{k} \mathbb
{E}\left[\Vert Y_t -\mathbf{v} \bar{\mathbf{y}}_t^{\intercal} \Vert_C\right].
\end{align}
The following parts are devoted to bound $M$.

$\mathbf{Step \ 1: \ Bound } \sum \limits _{t=1}^{k} \mathbb
{E}\left[\Vert \bar{\mathbf{y}}_t \Vert ^2\right]$  

It is deduced in a similar way to equation $\left(\ref{Theorem1_2}\right)$:
\begin{align}
-& \frac{\alpha}{n} \nabla F\left(\bar{\mathbf{x}}_{k}\right)^{\intercal} {Y_k}^{\intercal} \mathbf{u}\notag\\
=&- \frac{\alpha}{n} \nabla F\left(\bar{\mathbf{x}}_{k}\right)^{\intercal} \left({Y_k-\mathbf{v}\bar{\mathbf{y}}_k}^{\intercal}\right)^{\intercal} \mathbf{u}-\frac{\alpha}{n}\left(\mathbf{v}^{\intercal}\mathbf{u}\right) \Vert \bar{\mathbf{y}}_k \Vert^2 \notag\\
&- \frac{\alpha}{n} \left(\mathbf{v} ^{\intercal} \mathbf{u}\right) \left(\nabla F\left(\bar{\mathbf{x}}_k\right)-\bar{\mathbf{y}}_k\right)^{\intercal}\bar{\mathbf{y}}_k \notag\\
=&- \frac{\tilde{\alpha}}{\mathbf{v} ^{\intercal} \mathbf{u}} \nabla F\left(\bar{\mathbf{x}}_{k}\right)^{\intercal} \left({Y_k-\mathbf{v}\bar{\mathbf{y}}_k}^{\intercal}\right)^{\intercal} \mathbf{u}-\tilde{\alpha} \Vert \bar{\mathbf{y}}_k \Vert^2 \notag\\
&- \tilde{\alpha} \left(\frac{1}{n}\mathbf{1}^{\intercal} \left(\nabla F\left(\mathbf{1}{\bar{\mathbf{x}}_k}^{\intercal}\right)-\nabla F\left(X_k\right)\right)\right)^{\intercal}\bar{\mathbf{y}}_k + \tilde{\alpha} \frac{\gamma}{n} \mathbf{1}^{\intercal} E_k \bar{\mathbf{y}}_k \notag\\
\leq& \ \frac{\tilde{\alpha}}{\mathbf{v} ^{\intercal} \mathbf{u}} \nabla  \Vert F\left(\bar{\mathbf{x}}_{k}\right) \Vert \cdot \Vert {Y_k-\mathbf{v}\bar{\mathbf{y}}_k}^{\intercal} \Vert \cdot \Vert \mathbf{u} \Vert-\tilde{\alpha} \Vert \bar{\mathbf{y}}_k \Vert^2 \notag\\
& +\frac{L \tilde{\alpha}}{\sqrt{n}} \Vert \bar{\mathbf{y}}_k \Vert \cdot \Vert X_k - \mathbf{1}\bar{\mathbf{x}}_k ^{\intercal} \Vert + \tilde{\alpha} \frac{\gamma}{n} \mathbf{1}^{\intercal} E_k \bar{\mathbf{y}}_k \notag \\
\leq & \ L^2 n \tilde{\alpha} \Vert X_k - \mathbf{1}\bar{\mathbf{x}}_k ^{\intercal} \Vert^2 +\frac{2 \tilde{\alpha}}{\left(\mathbf{v}^{\intercal}\mathbf{u}\right)^2 } \Vert \mathbf{u} \Vert ^2 \cdot \Vert {Y_k-\mathbf{v}\bar{\mathbf{y}}_k}^{\intercal} \Vert ^2 \notag \\
& - \frac{\tilde{3\alpha}}{4} \Vert \bar{\mathbf{y}}_k \Vert^2 + \frac{\tilde{\alpha}}{8} \Vert \nabla F\left(\bar{\mathbf{x}}_k\right) \Vert^2 +\tilde{\alpha} \frac{\gamma}{n} \mathbf{1}^{\intercal} E_k \bar{\mathbf{y}}_k.
\label{Theorem1_4}
\end{align}
Combining $\left(\ref{Theorem1_1}\right)$, $\left(\ref{Theorem1_4}\right)$ and taking exception, we obtain that
\begin{align}
&\mathbb{E}\left[F\left(\bar{\mathbf{x}}_{k+1}\right)\right] \notag \\
\leq& \mathbb{E}\left[F\left(\bar{\mathbf{x}}_{k}\right)\right] +\frac{L \eta^2}{2 n^2} \Vert \mathbf{u} \Vert^2 \sigma_\varepsilon^2 + L^2n \tilde{\alpha} \cdot \mathbb{E}\left[\Vert X_k- \mathbf{1} \bar{\mathbf{x}}_k^{\intercal} \Vert^2\right] \notag \\
&+ \frac{\left(2+L\tilde{\alpha}\right )\tilde{\alpha}}{\left(\mathbf{v}^{\intercal}\mathbf{u}\right)^2}\Vert \mathbf{u}\Vert^2 \cdot \mathbb{E}\left[\Vert Y_k -\mathbf{v} \bar{\mathbf{y}}_k^{\intercal} \Vert\right] \notag \\ &+\left(L\tilde{\alpha}-\frac{3}{4}\right) \tilde{\alpha} \cdot \mathbb{E} \left[\Vert\bar{ \mathbf{y}}_k \Vert^2\right] + \frac{\tilde{\alpha}}{8} \mathbb{E} \left[\Vert \nabla F \left(\bar{\mathbf{x}}_k\right)  \Vert ^2\right].
\label{Theorem1_5}
\end{align}

Summing the equation $\left(\ref{Theorem1_5}\right)$ from $1$ to $k$, we can obtain that
\begin{align}
&\left(\frac{3}{4}-L\tilde{\alpha}\right)\sum \limits _{t=1}^k \mathbb{E} \left[\Vert\bar{ \mathbf{y}}_t \Vert^2\right] \notag \\
\leq &  \frac{\left[F\left(\bar{\mathbf{x}}_1\right)-F^{\star}\right]}{\tilde{\alpha}} + L^2 n  \sum \limits _{t=1}^k \mathbb{E} \left[\Vert X_k- \mathbf{1} \bar{\mathbf{x}}_t^{\intercal} \Vert^2\right] \notag \\
&+\frac{\left(2+L\tilde{\alpha}\right )\tilde{\alpha}}{\left(\mathbf{v}^{\intercal}\mathbf{u}\right)^2}\Vert \mathbf{u}\Vert^2 \cdot \sum \limits _{t=1}^k \mathbb{E}\left[\Vert Y_t -\mathbf{v} \bar{\mathbf{y}}_t^{\intercal} \Vert^2\right] \notag \\
&+\frac{1}{8} \sum \limits _{t=1}^k \mathbb{E} \left[\Vert \nabla F \left(\bar{\mathbf{x}}_t\right)  \Vert ^2\right] +k\frac{L \eta^2}{2 \tilde{\alpha} n^2} \Vert \mathbf{u} \Vert^2 \sigma_\varepsilon^2.
\label{Theorem1_6}
\end{align}

By letting $\frac{3}{4}-L\tilde{\alpha} > \frac{1}{4}$, i.e. $\alpha$ need satisfies
\begin{align}
\alpha < \frac{n}{2L \left(\mathbf{v}^{\intercal}\mathbf{u}\right)}. 
\label{alpha_1}
\end{align}
Thus, the equation $\left(\ref{Theorem1_6}\right)$ with Lemma 4 implies that
\begin{align}
&\sum \limits _{t=1}^k \mathbb{E} \left[\Vert\bar{ \mathbf{y}}_t \Vert^2\right] \notag \\
\leq &  \frac{4\left[F\left(\bar{\mathbf{x}}_1\right)-F^{\star}\right]}{\tilde{\alpha}} + 4L^2 n  \sum \limits _{t=1}^k \mathbb{E} \left[\Vert X_k- \mathbf{1} \bar{\mathbf{x}}_t^{\intercal} \Vert_R^2\right] \notag \\
&+\frac{10}{\left(\mathbf{v}^{\intercal}\mathbf{u}\right)^2}\Vert \mathbf{u}\Vert^2 \cdot \sum \limits _{t=1}^k \mathbb{E}\left[\Vert Y_t -\mathbf{v} \bar{\mathbf{y}}_t^{\intercal} \Vert_C^2\right] \notag \\
&+\frac{1}{2} \sum \limits _{t=1}^k \mathbb{E} \left[\Vert \nabla F \left(\bar{\mathbf{x}}_t\right)  \Vert ^2\right] +k\frac{2L \eta^2}{\tilde{\alpha}n^2} \Vert \mathbf{u} \Vert^2 \sigma_\varepsilon^2.
\label{tau_3}
\end{align}

$\mathbf{Step \ 2: \ Bound } \ \mathbb
{E}\left[\Vert X_{k+1}-\mathbf{1}\bar{\mathbf{x}}_k+1^{\intercal} \Vert_R ^2\right]$  

By Lemma 1 and $\left(\ref{definition_bar}\right)$,
\begin{align}
&X_{k+1}-\mathbf{1}\bar{\mathbf{x}}_{k+1}^{\intercal}=\left(I-\frac{\mathbf{1}\mathbf{u}^{\intercal}}{n}\right)\left(R_{\eta} X_k+\eta \Xi_k -\alpha Y_k \right) \notag \\
=&R_\eta X_k-\mathbf{1}\bar{\mathbf{x}}_k^{\intercal}-\alpha\left(I-\frac{\mathbf{1}\mathbf{u}^{\intercal}}{n}\right)Y_k+\eta \left(I-\frac{\mathbf{1}\mathbf{u}^{\intercal}}{n}\right) \notag \\
=&\left(R_\eta-\frac{\mathbf{1}\mathbf{u}^{\intercal}}{n}\right)\left(X_k-\mathbf{1}\bar{\mathbf{x}}_k^{\intercal}\right) -\alpha\left(I-\frac{\mathbf{1}\mathbf{u}^{\intercal}}{n}\right)Y_k \notag \\
&+\eta \left(I-\frac{\mathbf{1}\mathbf{u}^{\intercal}}{n}\right).
\label{Theorem1_7}
\end{align}
Combining $\left(\ref{Theorem1_7}\right)$ with Lemma 2 and 4 we can obtain that
\begin{align}
&\mathbb{E}\left[\Vert X_{k+1}-\mathbf{1}\bar{\mathbf{x}}_k^{\intercal}\Vert^2_R\right] \notag \\
=&\Vert R_\eta-\frac{\mathbf{1}\mathbf{u}^{\intercal}}{n} \Vert_R^2 \cdot \mathbb{E} \left[\Vert X_k-\mathbf{1}\bar{\mathbf{x}}_k^{\intercal}\Vert_R^2\right]  \notag \\
& -\alpha^2\Vert I-\frac{\mathbf{1}\mathbf{u}^{\intercal}}{n}\Vert_R^2 \cdot \mathbb{E} \left[\Vert Y_k \Vert^2_R\right] +\eta\Vert I-\frac{\mathbf{1}\mathbf{u}^{\intercal}}{n}\Vert_R^2 \cdot \mathbb{E} \left[\Vert \Xi_k \Vert _R^2\right] \notag \\
\leq&\ \eta^2 {\delta_{R,2} }^2 \Vert I-\frac{\mathbf{1}\mathbf{u}^{\intercal}}{n}  \Vert_R^2 \cdot \sigma_\varepsilon^2 + \frac{1+ {\tau_R}^2}{2}\cdot \mathbb{E} \left[\Vert X_k-\mathbf{1}\bar{\mathbf{x}}_k^{\intercal}\Vert_R^2\right] \notag \\
&+ \frac{2\alpha^2 \left(1+{\tau_R}^2\right)}{1-{\tau_R}^2} \delta_{R,2}^2 \Vert I-\frac{\mathbf{1}\mathbf{u}^{\intercal}}{n}\Vert_R^2 \cdot \mathbb{E} \left[\Vert Y_k - \mathbf{v} \bar{\mathbf{y}}_k^{\intercal} \Vert^2_C\right] \notag \\
&+    \frac{2\alpha^2 \left(1+{\tau_R}^2\right)}{1-{\tau_R}^2}\delta_{R,2}^2 \Vert \mathbf{v} \Vert^2 \Vert I-\frac{\mathbf{1}\mathbf{u}^{\intercal}}{n}\Vert_R^2 \cdot \mathbb{E} \left[\Vert \bar{\mathbf{y}}_k \Vert^2\right].
\label{Theorem1_8}
\end{align}

Define $\tau_u^2 \ := \ \frac{ 1+{\tau_R}^2}{1-{\tau_R}^2} \Vert I-\frac{\mathbf{1}\mathbf{u}^{\intercal}}{n}\Vert_R^2$ and we have $\frac{1+{\tau_R}^2}{x} < 1 < \frac{ 1+{\tau_R}^2}{1-{\tau_R}^2} $, it is easy to simplify equation $\left(\ref{Theorem1_8}\right)$ to the following:
\begin{align}
&\mathbb{E}\left[\Vert X_{k+1}-\mathbf{1}\bar{\mathbf{x}}_k^{\intercal}\Vert^2_R\right] \notag \\
\leq& \  \frac{1+ {\tau_R}^2}{2}\cdot \mathbb{E} \left[\Vert X_k-\mathbf{1}\bar{\mathbf{x}}_k^{\intercal}\Vert_R^2\right] \notag \\
&+ 2 \alpha^2 \tau_u^2 \delta_{R,2}^2 \cdot \mathbb{E} \left[\Vert Y_k - \mathbf{v} \bar{\mathbf{y}}_k^{\intercal} \Vert^2_C\right] \notag \\
&+ 2\alpha^2 \tau_u^2 \delta_{R,2}^2 \Vert \mathbf{v} \Vert^2 \cdot \mathbb{E} \left[\Vert \bar{\mathbf{y}}_k \Vert^2\right]+\eta^2 \tau_u^2 {\delta_{R,2} }^2 \cdot \sigma_\varepsilon^2.
\label{sigle_tau_1}
\end{align}

$\mathbf{Step \ 3: \ Bound } \ \mathbb
{E}\left[\Vert Y_{k+1}-\mathbf{v}\bar{\mathbf{y}}_{k+1}^{\intercal} \Vert_C ^2\right]$

By Lemma 1 and $\left(\ref{definition_bar}\right)$,
\begin{align}
&Y_{k+1}-\mathbf{v}\bar{\mathbf{y}}_{k+1}^{\intercal} \notag \\
=& \left(I- \frac{\mathbf{v}\mathbf{1}^{\intercal}}{n}\right) \left(C_\gamma Y_k+ \tilde{\nabla}F\left(X_k+1\right)-\tilde{\nabla}F\left(X_k\right)\right) \notag \\
=& \left(C_\gamma-\frac{\mathbf{v}\mathbf{1}^{\intercal}}{n}\right)\left(Y_{k}-\mathbf{v}\bar{\mathbf{y}}_{k}^{\intercal}\right) \notag \\
&+\left(I- \frac{\mathbf{v}\mathbf{1}^{\intercal}}{n}\right) \left( \tilde{\nabla}F\left(X_k+1\right)-\tilde{\nabla}F\left(X_k\right)\right).
\end{align}

Define $\tau_v^2 \ := \ \frac{ 1+{\tau_C}^2}{1-{\tau_C}^2} \Vert I-\frac{\mathbf{v}\mathbf{1}^{\intercal}}{n}\Vert_C^2$, similarly, we obtain that
\begin{align}
&\mathbb{E} \left[ \Vert Y_{k+1}-\mathbf{v}\bar{\mathbf{y}}_{k+1}^{\intercal} \Vert_C^2\right] \notag \\
\leq & \frac{1+{\tau_C}^2}{2} \cdot \mathbb{E} \left[ \Vert Y_{k}-\mathbf{v}\bar{\mathbf{y}}_{k}^{\intercal} \Vert_C^2\right] \notag \\
&+ {\tau_v}^2 \cdot \mathbb{E} \left[\Vert \tilde{\nabla}F\left(X_k+1\right)-\tilde{\nabla}F\left(X_k\right)\Vert_C^2\right].
\label{Theorem1_9}
\end{align}

It follows from \cite{sGT} that
\begin{align}
&\mathbb{E} \left[\Vert \tilde{\nabla}F\left(X_k+1\right)-\tilde{\nabla}F\left(X_k\right)\Vert_C^2\right] \notag \\
\leq & \delta_{C,2}^2 \cdot \left(\mathbb{E} \left[\Vert {\nabla}F\left(X_k+1\right)-{\nabla}F\left(X_k\right)\Vert_C^2\right] \right. \notag \\
& \qquad \qquad \qquad \qquad \left. + 2\gamma^2\left(1 + \alpha L\right)\cdot \sigma_\xi \right).
\label{Theorem1_10}
\end{align} 

Moreover, using the Lipschitz smoothness of the function $F$, we derive that
\begin{align}
&\mathbb{E} \left[\Vert {\nabla}F\left(X_k+1\right)-{\nabla}F\left(X_k\right)\Vert^2\right] \notag \\
\leq & \ L^2 \cdot \mathbb{E} \left[\Vert \left(R_\eta-I\right)X_{k}-\alpha Y_k+\eta \Xi_k \Vert^2  \right] \notag \\
= & \ L^2 \cdot \mathbb{E} \left[\Vert \left(R_\eta-I\right)\left(X_{k}-\mathbf{1}\bar{\mathbf{x}}_k^{\intercal}\right) \right. \notag \\
&   \qquad \qquad  \left. -\alpha \left(Y_k-\mathbf{v}\bar{\mathbf{y}}_k^{\intercal}+\alpha\mathbf{v}\bar{\mathbf{y}}_k^{\intercal}\right)\Vert^2  \right]+L^2 \eta^2 \cdot \sigma_\varepsilon \notag \\
\leq & \ L^2 \eta^2 \cdot \sigma_\varepsilon+ 3L^2\eta^2 \Vert R-I\Vert^2 \cdot  \mathbb{E}\left[\Vert X_{k}-\mathbf{1}\bar{\mathbf{x}}_k^{\intercal} \Vert^2 \right] \notag \\
&+3L^2 \alpha^2 \mathbb{E}\left[\Vert Y_k-\mathbf{v}\bar{\mathbf{y}}_k^{\intercal} \Vert^2\right]+  3L^2 \alpha^2 \Vert \mathbf{v}\Vert^2 \cdot \mathbb{E}\left[\Vert \bar{\mathbf{y}}_k^{\intercal} \Vert^2 \right]
\label{Theorem1_11}
\end{align}
where the last inequality follows the definition of $R_\eta$. Combining equations $\left(\ref{Theorem1_9}\right) - \left(\ref{Theorem1_11}\right)$, we have 
\begin{align}
&\mathbb{E} \left[ \Vert Y_{k+1}-\mathbf{v}\bar{\mathbf{y}}_{k+1}^{\intercal} \Vert_C^2\right] \notag \\
\leq & 3L^2 \eta^2 {\tau_v}^2 \delta_{C,2}^2 \Vert R-I \Vert^2 \cdot \mathbb{E}\left[\Vert X_{k}-\mathbf{1}\bar{\mathbf{x}}_k^{\intercal} \Vert^2 \right] \notag \\
&+ \left(3L^2 {\tau_v}^2 \delta_{C,2}^2\alpha^2 + \frac{1+{\tau_C}^2}{2}\right) \cdot \mathbb{E} \left[ \Vert Y_{k}-\mathbf{v}\bar{\mathbf{y}}_{k}^{\intercal} \Vert_C^2\right] \notag \\
&+ 3L^2 {\tau_v}^2 \delta_{C,2}^2 \Vert \mathbf{v} \Vert ^2 \alpha^2 \cdot \mathbb{E}\left[\Vert \bar{\mathbf{y}}_k^{\intercal} \Vert^2 \right] \notag \\
&+ L^2 {\tau_v}^2 \delta_{C,2}^2\eta^2 \cdot \sigma_\varepsilon^2 +2{\tau_v}^2 \delta_{C,2}^2 \gamma^2 \left(1+\alpha L\right) \cdot \sigma_\xi^2.
\label{sigle_tau_2}
\end{align}

$\mathbf{Step \ 4: \ Bound }  \mathbb
{E}\left[\Vert X_{t}-\mathbf{1}\bar{\mathbf{x}}_t^{\intercal} \Vert_R ^2\right] $ 

Combining $\left(\ref{sigle_tau_1}\right)$ and $\left(\ref{sigle_tau_2}\right)$ leads to the following matrix inequality
\begin{align}
&\underbrace{\begin{bmatrix}\Vert X_{k+1}-{\mathbf {1}}\bar{\mathbf {x}}_{k+1}^{\intercal}\Vert _R \\
	\Vert Y_{k+1}-\mathbf{v}\bar{\mathbf {y}}_{k+1}^{\intercal}\Vert _C\end{bmatrix}}_{\textstyle \triangleq \mathbf {z}_{k+1}}\preccurlyeq \underbrace{\begin{bmatrix}P_{11} & P_{12} \\
	P_{21} & P_{22} \end{bmatrix}}_{\textstyle \triangleq P} \underbrace{\begin{bmatrix}\Vert X_{k}-{\mathbf {1}}\bar{\mathbf {x}}_{k}^{\mathsf {T}}\Vert _{R}\\
	\Vert Y_{k}-\mathbf{v}\bar{\mathbf {y}}_{k}^{\mathsf {T}}\Vert _{C}\end{bmatrix}}_{\textstyle \triangleq \mathbf {z}_{k}}  +\underbrace{\begin{bmatrix} U_1 \\
	U_2 \end{bmatrix}}_{\textstyle \triangleq \mathbf {u}_{k}}
\label{matrix_inequality}
\end{align}
where $\preccurlyeq$ denotes the element-wise less than or equal sign and 
\begin{align*}
P_{11}=&\frac{1+ {\tau_R}^2}{2}, \ P_{12}=2 \alpha^2 \tau_u^2 \delta_{R,2}^2, \\
P_{21}=&3L^2 \eta^2 {\tau_v}^2 \delta_{C,2}^2 \Vert R-I \Vert^2, \\
P_{22}=&3L^2 {\tau_v}^2 \delta_{C,2}^2\alpha^2 + \frac{1+{\tau_C}^2}{2}, \\
U_1=& 2\alpha^2 \tau_u^2 \delta_{R,2}^2 \Vert \mathbf{v} \Vert^2 \cdot \mathbb{E} \left[\Vert \bar{\mathbf{y}}_k \Vert^2\right]+\eta^2 \tau_u^2 {\delta_{R,2} }^2 \cdot \sigma_\varepsilon^2, \\
U_2=&3L^2 {\tau_v}^2 \delta_{C,2}^2 \Vert \mathbf{v} \Vert ^2 \alpha^2 \cdot \mathbb{E}\left[\Vert \bar{\mathbf{y}}_k^{\intercal} \Vert^2 \right] \\
&+L^2 {\tau_v}^2 \delta_{C,2}^2\eta^2 \cdot \sigma_\varepsilon^2 +2{\tau_v}^2 \delta_{C,2}^2 \gamma^2 \left(1+\alpha L\right) \cdot \sigma_\xi^2.
\end{align*}

From Lemma 2, it is known that $\tau_R$, $\tau_C<1$ holds, and we have
\begin{align*}
\lim \limits _{\alpha \rightarrow 0} P=
\begin{bmatrix}
\frac{1+ {\tau_R}^2}{2} & 0 \\
P_{21} & \frac{1+ {\tau_C}^2}{2}
\end{bmatrix}.
\end{align*}

So, the spectral radius of the matrix $P$ in equation $\left(\ref{matrix_inequality}\right)$ is less than $1$ , i.e., $\rho\left(P\right)<1$ when $\alpha$ is sufficiently small. 

Let $\overline{\theta}$, $\underline{\theta}$ be the two eigenvalues of $P$. Without loss of generality, we assume that $\vert \underline{\theta} \vert < \vert \overline{\theta} \vert = \rho\left(P\right) \triangleq \theta <1$. Considering the eigendecomposition $P=T\Lambda T^{-1}$ with $\Lambda=\text{diag}\left(\underline{\theta},\overline{\theta}\right)$. With some tedious computations, we have 
\begin{align*}
\underline{\theta}=\frac{P_{11}+P_{22}-\psi}{2}, \ \overline{\theta}=\frac{P_{11}+P_{22}+\psi}{2}.
\end{align*}
where $\psi=\sqrt{\left(P_{11}-P_{22}\right)^2+2P_{12}P_{21}} \geq \vert P_{11}- P_{22} \vert \geq 0$.
Letting $\rho\left(P\right)=\overline{\theta}<1$, we obtain an upper bound of $\alpha$:
\begin{align}
\alpha < \left[ \frac{\left( 1-\tau _R^2 \right) \left( 1-\tau _C^2 \right)}{24L^2\tau _u^2\tau _v^2\delta _{R,2}^{2}\delta _{C,2}^{2} \eta^2 \Vert R-I\Vert^2+3L^2\tau _v^2\delta _{C,2}^{2}\left( 1-\tau _R^2 \right)} \right] ^{\frac{1}{2}}
\label{alpha_2}
\end{align}
In the similar way to \cite{PUSHPULL5}, we have that
\begin{align}
P^k=T\Lambda^k T^{-1} \preccurlyeq \theta^k 
\begin{bmatrix} 
1 & P_{12}\frac{k}{\theta} \\ P_{21}\frac{k}{\theta} & 1
\end{bmatrix}.
\end{align}

Matrix inequality $\left(\ref{matrix_inequality}\right)$ implies that 
\begin{align}
\mathbf{z}_k \preccurlyeq P^k \cdot \mathbf{z}_1 + \sum \limits_{t=1}^k P^{k-t}U_t.
\end{align}

Thus, we have
\begin{align}
\mathbb{E}\left[ \Vert X_k-1\mathbf{\bar{x}}_{k}^{\intercal}\Vert _{R}^{2} \right] \le P^k \cdot Z_1\left[ 1 \right] +\sum_{t=1}^k{P^{k-t}U_t\left[ 1 \right]}.
\label{Theorem1_12}
\end{align}
where $ P^k \cdot Z_1\left[ 1 \right]$, $U_t\left[ 1 \right]$ as shown follows denote the first row element of $P^k \cdot Z_1$, $U_t$, respectively:
\begin{align}
&P^kZ_1\left[ 1 \right] =\theta ^k\mathbb{E}\left[ \Vert  X_1-1\mathbf{\bar{x}}_{1}^{\intercal} \Vert_{R}^{2} \right] +k\theta ^k\frac{P^{12}}{\theta}\mathbb{E}\left[ \Vert Y_1-\mathbf{v\bar{y}}_{1}^{\intercal}\Vert _{C}^{2} \right] \label{Theorem1_13}\\
&P^{k-t}U_t\left[ 1 \right]  \notag \\
&=\theta ^{k-t}\left( 2\alpha ^2\tau _{u}^{2}\delta _{R,2}^{2} \Vert \mathbf{v }\Vert^2 \mathbb
{E}\left[ \Vert \mathbf{\bar{y}}_t\Vert ^2 \right] +\eta ^2\tau _{u}^{2}\delta _{R,2}^{2}\sigma _{\xi}^{2} \right) \notag \\
&\ \ +P^{12}\left( k-t \right) \theta ^{k-t-1}\tau _{v}^{2}\delta _{C,2}^{2}\left( 3L^2 \Vert \mathbf{v} \Vert ^2\alpha ^2\cdot \mathbb{E}\left[ \Vert \mathbf{\bar{y}}_t \Vert ^2 \right] \right. \notag \\ 
&\ \ \left. +L^2\eta ^2\sigma _{\xi}^{2}+2\gamma ^2\left( 1+\alpha L \right) \sigma _{\varepsilon}^{2} \right) .
\label{Theorem1_14}
\end{align}

We have $\sum \limits_{t=1}^k\theta^t <\sum \limits_{t=1}^{\infty}\theta^t \leq \frac{1}{1-\theta}$, $\sum \limits_{t=1}^k t\theta^t <\sum \limits_{t=1}^{\infty} t\theta^t \leq \frac{1}{\left(1-\theta\right)^2}$, since $\theta<1$ holds when $\alpha$ satisfies specific conditons. It follows from equation $\left(\ref{Theorem1_13}\right)$ that 
\begin{align}
&\sum \limits_{t=1}^k{P^tZ_1\left[ 1 \right]}\notag \\
\leq & \frac{1}{1-\theta}\mathbb{E}\left[ \Vert X_1-1\mathbf{\bar{x}}_{1}^{\intercal}\Vert _{R}^{2} \right] +\frac{2\alpha ^2\tau _{u}^{2}\delta _{R,2}^{2}}{\theta \left( 1-\theta \right) ^2}\mathbb{E}\left[ \Vert Y_1-\mathbf{v\bar{y}}_{1}^{\intercal}\Vert _{C}^{2} \right] .
\label{Theorem1_15}
\end{align}

The equation $\left(\ref{Theorem1_14}\right)$ combined with the Lemmma 4.5 in \cite{sGT1} yeilds
\begin{align}
&\sum \limits_{t=1}^k{\sum \limits_{i=1}^t{P^{t-i}U_i\left[ 1 \right]}} \notag \\
\leq & \frac{1}{1-\theta}\sum_{t=1}^k{\left( 2\alpha ^2\tau _{u}^{2}\delta _{R,2}^{2} \Vert \mathbf{v}\Vert ^2 \cdot \mathbb{E}\left[ \Vert \mathbf{\bar{y}}_t\Vert ^2 \right] +\eta ^2\tau _{u}^{2}\delta _{R,2}^{2}\sigma _{\xi}^{2} \right)} \notag \\
&+\frac{2\alpha ^2\tau _{u}^{2}\delta _{R,2}^{2}}{\left( 1-\theta \right) ^2}\sum_{t=1}^k \left( 3L^2\tau _{v}^{2}\delta _{C,2}^{2} \Vert \mathbf{v}\Vert ^2\alpha ^2\cdot \mathbb{E}\left[ \Vert \mathbf{\bar{y}}_t\Vert ^2 \right] \right.\notag \\
&\left. +L^2\tau _{v}^{2}\eta ^2\delta _{C,2}^{2}\sigma _{\xi}^{2}+2\tau _{v}^{2}\delta _{C,2}^{2}\gamma ^2\left( 1+\alpha L \right) \sigma _{\varepsilon}^{2} \right) \notag \\
=&\frac{2\tau _{u}^{2}\Vert \mathbf{v}\Vert ^2\delta _{R,2}^{2}}{1-\theta}\alpha ^2\left( 1+\frac{3L^2\tau _{v}^{2}\delta _{C,2}^{2}}{1-\theta}\alpha ^2 \right) \sum_{t=1}^k{ \mathbb{E}\left[ \Vert \mathbf{\bar{y}}_t\Vert ^2 \right]} \notag \\
&+\frac{\eta ^2\tau _{u}^{2}\delta _{R,2}^{2}}{1-\theta}\left( 1+\frac{2L^2\tau _{v}^{2}\delta _{C,2}^{2}}{1-\theta}\alpha ^2 \right) k\sigma _{\xi}^{2} \notag \\
&+\frac{4\tau _{u}^{2}\tau _{v}^{2}\gamma ^2\left( 1+\alpha L \right) \delta _{R,2}^{2}\delta _{C,2}^{2}}{\left( 1-\theta \right) ^2}\alpha ^2k\sigma _{\varepsilon}^{2}.
\label{Theorem1_16}
\end{align}

Define a parameter $\tau_\alpha \triangleq \frac{n}{2L\left(\mathbf{v}^{\intercal}\mathbf{u}\right)}> \alpha$ as discussed in $\mathbf{Step \ 1}$. It follows from $\left(\ref{Theorem1_12}\right)$, $\left(\ref{Theorem1_15}\right)$ and $\left(\ref{Theorem1_16}\right)$ that
\begin{align}
&\sum \limits_{t=1}^k \mathbb
{E}\left[\Vert X_{t}-\mathbf{1}\bar{\mathbf{x}}_t^{\intercal} \Vert_R ^2\right]= \sum \limits_{t=1}^k{P^tZ_1\left[ 1 \right]} + \sum \limits_{t=1}^k{\sum \limits_{i=1}^t{P^{t-i}U_i\left[ 1 \right]}} \notag \\
&\leq  g_1+g_2\alpha^2 \cdot \sum\limits_{t=1}^k \mathbb{E}\left[\Vert \bar{\mathbf {y}}_t^2 \Vert\right]+\eta^2 g_3 k \cdot \sigma_\xi^2+ \gamma^2 g_4 k \cdot \sigma_\varepsilon^2.
\label{tau_1}
\end{align}
where $g1$, $g2$, $g3$, $g4$ are time independent constants as follows
\begin{align*}
&g_1=\frac{1}{1-\theta}\mathbb{E}\left[ \Vert X_1-1\mathbf{\bar{x}}_{1}^{T}\Vert _{R}^{2} \right] +\frac{2\tau _{u}^{2}\tau _{\alpha}^{2}\delta _{R,2}^{2}}{\theta \left( 1-\theta \right) ^2}\mathbb{E}\left[ \Vert Y_1-\mathbf{v\bar{y}}_{1}^{T}\Vert _{C}^{2} \right], \\
&g_2=\frac{2\tau _{u}^{2}\Vert \mathbf{v} \Vert^2\delta _{R,2}^{2}}{1-\theta}\left( 1+\frac{3L^2\tau _{v}^{2}\delta _{C,2}^{2}}{1-\theta}\tau _{\alpha}^{2} \right), \\
&g_3= \frac{\tau _{u}^{2}\delta _{R,2}^{2}}{1-\theta}\left( 1+\frac{2L^2\tau _{v}^{2}\delta _{C,2}^{2}}{1-\theta}\tau _{\alpha}^2 \right),\\
&g_4=\frac{4\tau _{u}^{2}\tau _{v}^{2}\left( 1+\tau _{\alpha}L \right) \tau _{\alpha}^{2}\delta _{R,2}^{2}\delta _{C,2}^{2}}{\left( 1-\theta \right) ^2}.
\end{align*}

$\mathbf{Step \ 5: \ Bound } \sum \limits_{t=1}^k \mathbb
{E}\left[\Vert Y_{t}-\mathbf{v}\bar{\mathbf{y}}_{t}^{\intercal} \Vert_C ^2\right]$

Similarly, $\sum \limits_{t=1}^k \mathbb
{E}\left[\Vert Y_{t}-\mathbf{v}\bar{\mathbf{y}}_{t}^{\intercal} \Vert_C ^2\right]$ can be upper bounded: 
\begin{align}
&\sum \limits_{t=1}^k \mathbb
{E}\left[\Vert Y_{t}-\mathbf{v}\bar{\mathbf{y}}_{t}^{\intercal} \Vert_C ^2\right]= \sum \limits_{t=1}^k{P^tZ_1\left[ 2 \right]} + \sum \limits_{t=1}^k{\sum \limits_{i=1}^t{P^{t-i}U_i\left[2 \right]}} \notag \\
&\leq  h_1+h_2\alpha^2 \cdot \sum\limits_{t=1}^k \mathbb{E}\left[\Vert \bar{\mathbf {y}}_t^2 \Vert\right]+\eta^2 h_3 k \cdot \sigma_\xi^2+ \gamma^2 h_4 k \cdot \sigma_\varepsilon^2
\label{tau_2}
\end{align}
where $h_1$, $h_2$, $h_3$, $h_4$ are time independent constants as follows
\begin{align*}
h_1=&\frac{3L^2\tau _{v}^{2}\delta _{C,2}^{2} \eta^2 \Vert R-I\Vert ^2}{\theta \left( 1-\theta \right) ^2}\mathbb{E}\left[\Vert X_1-1\mathbf{\bar{x}}_{1}^{\intercal}\Vert _{R}^{2} \right]\\ &+\frac{1}{1-\theta}\mathbb{E}\left[ \Vert Y_1-\mathbf{v\bar{y}}_{1}^{\intercal}\Vert _{C}^{2} \right], \\
h_2=&\frac{3L^2\tau _{v}^{2}\Vert \mathbf{v}\Vert ^2\delta _{C,2}^{2}}{1-\theta}\left( 1+\frac{2\tau _{u}^{2} \eta^2 \Vert R-I\Vert ^2\delta _{R,2}^{2}}{1-\theta} \right), \\
h_3=& \frac{L^2\tau _{v}^{2}\delta _{C,2}^{2}}{1-\theta}\left( 1+\frac{3\tau _{u}^{2} \eta^2 \Vert R-I\Vert ^2\delta _{R,2}^{2}}{1-\theta} \right),\\
h_4=&\frac{2\tau _{v}^{2}\left( 1+\tau _{\alpha}L \right) \delta _{C,2}^{2}}{1-\theta}.
\end{align*}

$\mathbf{Step \ 6: \ Bound }\ M$

Let $\tau_1=\mathbb
{E}\left[\Vert X_{t}-\mathbf{1}\bar{\mathbf{x}}_t^{\intercal} \Vert_R ^2\right]$, $\tau_2=\sum \limits_{t=1}^k \mathbb{E}\left[\Vert Y_{t}-\mathbf{v}\bar{\mathbf{y}}_{t}^{\intercal} \Vert_C ^2\right]$, $\tau_3=\mathbb {E}\left[\Vert Y_{k+1}-\mathbf{v}\bar{\mathbf{y}}_{k+1}^{\intercal} \Vert_C ^2\right]$. We assume that the term $M$ is upper bounded by $p^{\star}$, i.e., $M \leq p^{\star}$. Combining equations $\left(\ref{tau_3}\right)$, $\left(\ref{tau_1}\right)$ and $\left(\ref{tau_2}\right)$, obtaining $p^{\star}$ can be equivalent to calculate the optimal value following linear programming problem:
\begin{align}
\max \limits_{\tau_{1},\tau_{2},\tau_{3}>0} \ & c_1\tau _1+c_2\tau _2+c_3\tau _3 \notag \\
s.t. \quad \ \ &\tau _1\leq a_1\tau _3+b_1 \notag \\
&\tau _2\le a_2\tau _3+b_2 \notag \\
&\tau _3\le a_3\tau _1+a_4\tau _2+b_3 
\end{align}
where parameters are all positive as shown follows 
\begin{align*}
c_1=&\frac{11}{3}L^2n\ \ \ c_2=\frac{4\left( 2+L\tilde{\alpha} \right)}{3\left( \mathbf{v}^T\mathbf{u} \right) ^2} \Vert \mathbf{u} \Vert ^2\ \ \ \ c_3=\frac{4}{3}L\tilde{\alpha} \\
a_1=&g_2\alpha ^2\ \ \ \ \ \ b_1=g_1+\eta ^2g_3k\sigma _{\xi}^{2}+\gamma ^2g_4k\sigma _{\varepsilon}^{2} \\
a_2=&h_2\alpha ^2\,\,\,\,\,\,\,\,\ \ b_2=h_1+\eta ^2h_3k\sigma _{\xi}^{2}+\gamma ^2h_4k\sigma _{\varepsilon}^{2} \\
a_3=&4L^2n\,\,\,\,\,\,\,\,\ \ a_4=\frac{10}{\left( \mathbf{v}^T\mathbf{u} \right) ^2} \Vert \mathbf{u} \Vert ^2 \\
b_3=&\frac{4\left[ F\left( \mathbf{\bar{x}}_1 \right) -F^{\star} \right]}{\tilde{\alpha}}+\frac{1}{2}\sum_{t=1}^k{ \mathbb{E}\left[  \Vert \nabla F\left( \mathbf{\bar{x}}_k \right)  \Vert ^2 \right]} \\
&+\frac{2L\eta ^2}{\tilde{\alpha}n^2} \Vert \mathbf{u} \Vert ^2k\sigma _{\xi}^{2}.
\end{align*}

Letting $1-a_2a_4-a_1a_3>0$, stepsize $\alpha$ should satisfies
\begin{align}
\alpha <\left[ \frac{\left( \mathbf{v}^{\intercal}\mathbf{u} \right) ^2}{4L^2n\left( \mathbf{v}^{\intercal}\mathbf{u} \right) ^2g_2+10\Vert \mathbf{u}\Vert ^2h_2} \right] ^{\frac{1}{2}}.
\label{alpha_3}
\end{align}
If $1-a_2a_4-a_1a_3>0$ holds, it can be checked that the optimal solution $\left(\tau_1^{\star}, \ \tau_2^{\star}, \ \tau_3^{\star}\right)$ can be obtained when all constraints are active, i.e. $\tau _{1}^{\star}=a_1\tau _{3}^{\star}+b_1$, $\tau _{2}^{\star}=a_2\tau _{3}^{\star}+b_2$ and $\tau _{3}^{\star}=a_3\tau _{1}^{\star}+a_4\tau _{2}^{*}+b_3$. By explicitly solving the problem, $p^{\star}$ is derived
\begin{align}
p^{\star} =& c_1\tau _{1}^{\star}+c_2\tau _{2}^{\star}+c_3\tau _{3}^{\star} \notag \\
=&c_1\frac{\left[ \left( 1-a_2a_4 \right) +a_2a_3+a_3 \right] b_1}{1-a_2a_4-a_1a_3}+c_3\frac{\left( a_1+a_2+1 \right) b_3}{1-a_2a_4-a_1a_3} \notag \\
&+c_2\frac{\left[ a_1a_4+\left( 1-a_1a_3 \right) +a_4 \right] b_2}{1-a_2a_4-a_1a_3} \notag \\
=& \left[1+\frac{a_1+a_2+1}{1-a_2a_4-a_1a_3}a_3\right]c_1b_1+ \frac{a_1+a_2+1}{1-a_2a_4-a_1a_3} c_3b_3\notag \\
&+\left[1+\frac{a_1+a_2+1}{1-a_2a_4-a_1a_3}a_4\right]c_2b_2.
\label{upperbound_p_1}
\end{align}

When $\alpha$ is sufficiently small, $c_3<\frac{1}{4}$ and $1-\frac{1}{2}\frac{a_1+a_2+1}{1-a_2a_4-a_1a_3}\geq\frac{3}{4}$. By tedious calculation, we have the following results to bound $\alpha$:
\begin{align}
&\alpha< \min \{\frac{3n}{16L\left( \mathbf{v}^{\intercal}\mathbf{u} \right)}\} \label{alpha_4},\\
&\alpha<\left[ \frac{1}{\left( 1+8L^2n \right) g_2+\left( 1+\frac{20}{\left( \mathbf{v}^T\mathbf{u} \right) ^2}\Vert \mathbf{u} \Vert^2 \right) h_2} \right] ^{\frac{1}{2}}. \label{alpha_5}
\end{align} 

When above upper bound of $\alpha$ are satisfied, it follows equation $\left( \ref{upperbound_p_1}\right)$ that
\begin{align}
p^{\star}\leq& \left( 1+2a_3 \right) c_1b_1+\left( 1+2a_4 \right) c_2b_2+\frac{1}{2}b_3 \notag \\
\leq& \left( G_1+\eta ^2G_3k\sigma _{\xi}^{2}+\gamma ^2G_4k\sigma _{\varepsilon}^{2} \right) \notag \\
&+\left( H_1+\eta ^2H_3k\sigma _{\xi}^{2}+\gamma ^2H_4k\sigma _{\varepsilon}^{2} \right) +\frac{2\left[ F\left( \mathbf{\bar{x}}_1 \right) -F^* \right]}{\tilde{\alpha}} \notag \\
&+\frac{1}{4}\sum_{t=1}^k{E\left[ \Vert \nabla F\left( \mathbf{\bar{x}}_t \right) \Vert ^2 \right]}+\frac{L\eta ^2}{\tilde{\alpha}n^2}\Vert \mathbf{u}\Vert ^2k\sigma _{\xi}^{2} .
\label{upperbound_p_2}
\end{align}
where $G_i, H_i,i=1,3,4$ (as shown below) are positive constants , which are independent of the iteration times, $G_i=\left[\frac{11}{3}L^2n\left(1+8L^2n\right)\right]g_i$, $H_i=\left[ \frac{35\Vert \mathbf{u}\Vert ^2}{12\left( \mathbf{v}^{\intercal}\mathbf{u} \right) ^2}\left( 1+\frac{20}{\left( \mathbf{v}^{\intercal}\mathbf{u} \right) ^2}\Vert \mathbf{u}\Vert ^2 \right) \right] h_i$.

Equation $\left(\ref{upperbound_p_2}\right)$ provide an upper bound for $M$. Combining equations $\left(\ref{target_inequality}\right)$ and $\left(\ref{upperbound_p_2}\right)$ and divide both side by $k$, we can obtain the following result:
\begin{align}
M\left(k\right)
\leq &\frac{10n\left[ F\left( \mathbf{\bar{x}}_1 \right) -F^* \right]}{3\left( \mathbf{v}^T\mathbf{u} \right) \alpha}\frac{1}{k}+\left( G_1+H_1 \right) \frac{1}{k} \notag \\
&+\left( G_3+H_3+\frac{5L}{3\left( \mathbf{v}^{\intercal}\mathbf{u} \right) \alpha n}\Vert \mathbf{u}\Vert^2 \right) \eta ^2\sigma _{\xi}^{2} \notag \\
&+\left( G_4+H_4 \right) \gamma ^2\sigma _{\varepsilon}^{2}.
\label{theorem1_final}
\end{align}

Thus, when the stepsize satisfies equations $\left(\ref{alpha_2}\right)$, $\left(\ref{alpha_3}\right)$, $\left(\ref{alpha_4}\right)$, $\left(\ref{alpha_5}\right)$, the above relation holds. We get the result consistent with the Theorem 1, the proof is completed.
\subsection{Proof of Theorem 3:}
When $\mathcal{W}=\mathbb{R}^m$ and cost functions of every nodes $f_i$ is differentiable, it follows equation $\left(12\right)$ that $\ \nabla f_i\left( \mathbf{w}_{i,k} \right) =-\mathbf{x}_{i,k}$, $\mathbf{w}_{i,k}=-\nabla F_i\left( \mathbf{x}_{i,k} \right) +\mathbf{d}_i$. Thus, we have 
\begin{align}
&\sum_{i=1}^n{\Vert \nabla f_i\left( \mathbf{w}_{i,k} \right) -\frac{1}{n}\sum_{j=1}^n{\nabla f_j\left( \mathbf{w}_{j,k} \right)}\Vert ^2} \notag \\
=&\sum_{i=1}^n{\Vert -\mathbf{x}_{i,k}-\frac{1}{n}\sum_{j=1}^n{\left( -\mathbf{x}_{j,k}\right)}\Vert ^2}=\Vert \left( I-\frac{1}{n}\mathbf{1}\mathbf{1}^{\intercal} \right) X_k\Vert ^2 \notag \\
\leq& \ 2\cdot \Vert \left( I-\frac{1}{n}\mathbf{1}\mathbf{u}^T \right) \left( X_k-\mathbf{1}\mathbf{\bar{x}}_{k}^{\intercal} \right)\Vert ^2 \notag \\
&+2 \cdot \Vert \left( \frac{1}{n}\mathbf{1}\mathbf{u}^T-\frac{1}{n}\mathbf{1}\mathbf{1}^{\intercal} \right) \left( X_k-\mathbf{1}\mathbf{\bar{x}}_{k}^{\intercal} \right)\Vert ^2 \notag \\
\leq & 2\left(1+ \Vert \frac{1}{n}\mathbf{1}\mathbf{u}^{\intercal}-\frac{1}{n}\mathbf{1}\mathbf{1}^{\intercal} \Vert^2 \right)\Vert X_k-1\mathbf{\bar{x}}_{k}^{\intercal}\Vert ^2 \notag \\
\leq & 2n \cdot \Vert X_k-1\mathbf{\bar{x}}_{k}^{\intercal}\Vert ^2.
\label{Proof3_1}
\end{align}
where the first inequality follows $\Vert X + Y \Vert^2 \leq 2\Vert X \Vert^2+2\Vert Y \Vert^2$ and the last inequality uses the relation $\Vert \frac{1}{n}\mathbf{1}\mathbf{u}^{\intercal}-\frac{1}{n}\mathbf{1}\mathbf{1}^{\intercal} \Vert^2 \leq n-1$

Moreover, it follows $\mathbf{w}_{i,k}=-\nabla F_i\left( \mathbf{x}_{i,k} \right) +\mathbf{d}_i$ that
\begin{align}
&\Vert \sum_{i=1}^n{\mathbf{w}_{i,k}}-D\Vert ^2 \notag \\
=&\Vert \sum_{i=1}^n{\left[ -\nabla F_i\left( \mathbf{x}_{i,k}\ \right) +\mathbf{d}_{i} \right]}-D\Vert ^2=\Vert \sum_{i=1}^n{\nabla F_i\left( \mathbf{x}_{i,k} \right)}\Vert ^2 \notag \\
=& \Vert \nabla F\left( \mathbf{\bar{x}}_k \right) +\sum_{i=1}^n{\left[ \nabla F_i\left( \mathbf{x}_{i,k} \right) -\nabla F_i\left( \mathbf{\bar{x}}_k \right) \right]}\Vert ^2 \notag \\
\leq& 2\Vert \nabla F\left( \mathbf{\bar{x}}_k \right) \Vert ^2+2n\sum_{i=1}^n{\Vert \nabla F_i\left( \mathbf{x}_{i,k} \right) -\nabla F_i\left( \mathbf{\bar{x}}_k \right) \Vert ^2} \notag \\
\leq & 2\Vert \nabla F\left( \mathbf{\bar{x}}_k \right) \Vert ^2+\frac{2n}{\mu^2}\Vert X_k-\mathbf{1}\mathbf{\bar{x}}_{k}^{\intercal}\Vert ^2
\label{Proof3_2}
\end{align}
where the first inequality uses the Cauchy-Schwarz inequality and the last inequality uses the $L$-Lipschutz Smooth $\left(L=\frac{1}{\mu}\right)$ of the function $F_i$.

Combining equations $\left(\ref{Proof3_1}\right)$ and $\left(\ref{Proof3_2}\right)$, we have
\begin{align}
N\left(k\right) \leq  2 \mathbb{E}\left[ \Vert \nabla F\left( \mathbf{\bar{x}}_k \right) \Vert ^2 \right] +2n\left( 1+L^2 \right) \mathbb{E}\left[ \Vert X_k-1\mathbf{\bar{x}}_{k}^{\intercal}\Vert ^2 \right].
\end{align}
Adjusting the coefficient in $\left(\ref{target_inequality}\right)$, we can obtain a new relation:
\begin{align}
&\sum_{t=1}^k{ \mathbb{E}\left[ \Vert \nabla F\left( \mathbf{\bar{x}}_t \right) \Vert ^2 \right] +\frac{3}{4}n\left( 1+L^2 \right) \cdot \mathbb{E}\left[ \Vert X_t-1\mathbf{\bar{x}}_{t}^{\intercal}\Vert^2 \right]} \notag \\
\leq & \frac{4\left[ F\left( \mathbf{\bar{x}}_1 \right) -F^{\star} \right]}{3\tilde{\alpha}}+\frac{2L\eta ^2}{3n^2\tilde{\alpha}} \Vert \mathbf{u}\Vert ^2k\sigma _{\xi}^{2}\ +\frac{4}{3}L\tilde{\alpha}\cdot \sum_{t=1}^k{\mathbb{E}\left[ \Vert \mathbf{\bar{y}}_t\Vert ^2 \right]} \notag \\
& \ +\left( \frac{41}{12}L^2+\frac{3}{4} \right) n\cdot \sum_{t=1}^k{\mathbb{E}\left[ \Vert X_t-1\mathbf{\bar{x}}_{t}^{\intercal}\Vert _{R}^{2} \right]} \notag \\
& \ +\frac{4\left( 2+L\tilde{\alpha} \right)}{3\left( \mathbf{v}^T\mathbf{u} \right) ^2}\Vert \mathbf{u}\Vert ^2\cdot \sum_{t=1}^k{\mathbb{E}\left[ \Vert Y_t-\mathbf{v\bar{y}}_{t}^{T}\Vert _{C}^{2} \right]}.
\label{target_inequality2}
\end{align}
We define the sum of the last three terms in equation $\left(\ref{target_inequality2}\right)$ as $M'$ and bound $M'$ in the similar way to the proof of Theorem 1. We can calculate its upper bound $p'^{\star}$ as follows:
\begin{align}
p'^{\star}\leq& \left( 1+2a_3 \right) c_1b_1+\left( 1+2a_4 \right) c_2b_2+\frac{1}{2}b_3 \notag \\
\leq & \left( G_{1}^{'}+\eta ^2G_{3}^{'}k\sigma _{\xi}^{2}+\gamma ^2G_{4}^{'}k\sigma _{\varepsilon}^{2} \right) \notag \\
& \ +\left( H'_1+\eta ^2H'_3k\sigma _{\xi}^{2}+\gamma ^2H'_4k\sigma _{\varepsilon}^{2} \right)  +\frac{2\left[ F\left( \mathbf{\bar{x}}_1 \right) -F^* \right]}{\tilde{\alpha}} \notag \\
& \ +\frac{1}{4}\sum_{t=1}^k{E\left[ \Vert \nabla F\left( \mathbf{\bar{x}}_t \right) \Vert ^2 \right]}+\frac{L\eta ^2}{\tilde{\alpha}n^2}\Vert \mathbf{u} \Vert ^2k\sigma _{\xi}^{2}
\label{bound_M'}
\end{align}
where $G'_i, H'_i,i=1,3,4$ (as shown below) are positive constants , which are independent of the iteration times, $G'_i=\left[ \left( \frac{41}{12}L^2+\frac{3}{4} \right) \left( 1+8L^2n \right) \right] g_i$, $H'_i=\left[ \frac{35\Vert \mathbf{u}\Vert ^2}{12\left( \mathbf{v}^{\intercal}\mathbf{u} \right) ^2}\left( 1+\frac{20}{\left( \mathbf{v}^{\intercal}\mathbf{u} \right) ^2}\Vert \mathbf{u}\Vert ^2 \right) \right] h_i$. Note that $g_i$, $h_i$ are defined in the proof of Theorem 1.

Moreover, it follows equations $\left(\ref{target_inequality2}\right)$ and $\left(\ref{bound_M'}\right)$ that
\begin{align}
N\left(k\right) \leq& \frac{40n\left[ F\left( \mathbf{\bar{x}}_1 \right) -F^* \right]}{9\left( \mathbf{v}^T\mathbf{u} \right) \alpha}\frac{1}{k}+\frac{4}{3}\left( G'_1+H'_1 \right) \frac{1}{k} \notag \\
&+\frac{4}{3}\left( G'_3+H'_3+\frac{5L}{3\left( \mathbf{v}^{\intercal}\mathbf{u} \right) \alpha n}\Vert \mathbf{u}\Vert^2 \right) \eta ^2\sigma _{\xi}^{2} \notag \\
&+\frac{4}{3}\left( G'_4+H'_4 \right) \gamma ^2\sigma _{\varepsilon}^{2}.
\end{align}

Thus, we get the result consistent with the Theorem 3, the proof is completed.
\end{document}